\documentclass[useAMS,usenatbib]{mn2e}

\usepackage[]{epsfig}
\usepackage[T1]{fontenc}
\usepackage{aecompl}
\usepackage{amsmath}
\usepackage{amssymb}
\usepackage{myaasmacros}
\usepackage[]{graphicx}
\DeclareGraphicsExtensions{.jpg,.png,.pdf,.eps,.ps}
\def\LCDM{$\Lambda\mbox{CDM}$  }

\def\Gadget2{\rm{\textsc{Gadget\thinspace 2}\ }}
\def\kms{{\rm\thinspace km\thinspace s}^{-1}}

\def\kpc{{\rm\thinspace kpc}}

\def\Msun{\hbox{$\mathrm{\thinspace M_{\odot}}$}}

\def\Myr{{\rm\thinspace Myr}}

\title[BHs in galaxy simulations - the regularized tree code rVINE]{Dynamical evolution of massive black holes in
galactic-scale N-body simulations -- introducing the regularized
tree code ``rVINE''}
\author[S. J. Karl, S. J. Aarseth,
  T. Naab, M. G. Haehnelt, R. Spurzem]{Simon
    J. Karl$^1$\thanks{skarl@ast.cam.ac.uk}, Sverre
    J. Aarseth$^2$, Thorsten Naab$^3$, Martin G. Haehnelt$^1$,
    \newauthor{and Rainer Spurzem$^4$$^,$$^5$$^,$$^6$}\\
$^1$ Institute of Astronomy and Kavli Institute for Cosmology,
Madingley Road, Cambridge CB3 0HA, United Kingdom;\\
$^2$ Institute of Astronomy, University of Cambridge,
Madingley Road, Cambridge CB3 0HA, United Kingdom;\\
$^3$ Max-Planck-Institut f\"ur Astrophysik,
  Karl-Schwarzschild-Str. 1, D-85741 Garching bei M\"unchen,
  Germany;\\
$^4$ National Astronomical Observatories of China, Chinese Academy of
Sciences, 20A Datun Rd., Chaoyang District, 100012, Beijing, China;\\
$^5$ Astronomisches Rechen-Institut, Zentrum f\"ur Astronomie,
University of Heidelberg, M\"onchhofstrasse 12-14, 69120, Heidelberg,
Germany;\\
$^6$ Kavli Institute for Astronomy \& Astrophysics, Yi He Yuan Lu 5,
Hai Dian Qu, 100871 Bejing, China}

\begin{document}

\pagerange{\pageref{firstpage}--\pageref{lastpage}} \pubyear{2012}

\maketitle

\label{firstpage}

\begin{abstract}
\label{abstract}
We present a hybrid code combining the OpenMP-parallel tree
code VINE with an algorithmic chain regularization scheme. The new code,
called ``rVINE'', aims to significantly improve the accuracy of close
encounters of massive bodies with supermassive black holes in
galaxy-scale numerical simulations.
We demonstrate the capabilities of the code by studying two test
problems, the sinking of a single massive black hole to the centre of
a gas-free galaxy due to dynamical friction and the hardening of a
supermassive black hole binary due to close stellar encounters. We show that
results obtained with rVINE compare well with NBODY7 
for problems with particle numbers that can be simulated with
NBODY7. In particular, in both NBODY7 and rVINE we find a clear
N-dependence of the binary hardening rate, a low binary
eccentricity and moderate eccentricity evolution, as well as the
conversion of the galaxy's inner density profile from a cusp to a a
core via the ejection of stars at high velocity.
The much larger number of particles that can be handled by rVINE will
open up exciting opportunities to model stellar dynamics close to SMBHs
much more accurately in a realistic galactic
context. This will help to remedy the inherent limitations of commonly
used tree solvers to follow the correct
dynamical evolution of black holes in galaxy scale simulations. 
\end{abstract}


\section{Introduction}
\label{Intro}

The presence of supermassive black holes \citep[SMBHs;
e.g.][]{1984ARA&A..22..471R} with masses of $10^6 \Msun \lesssim
M_{\rm{BH}} \lesssim 10^{10} \Msun$ hosted in the central regions of
virtually all massive spheroids in the nearby Universe,
including the Galactic bulge, is now firmly established
\citep{1998Natur.395A..14R, 2013ARA&A..51..511K}. SMBHs must have also
been present at much earlier phases of our Universe, powering active
galactic nuclei (AGNs) and quasars from only a few hundred million
years after the Big Bang throughout cosmic history
\citep[e.g.][]{1969Natur.223..690L, 2001AJ....122.2833F,
  2011ApJ...741...91C, 2011Natur.474..616M}. 

We now think of SMBHs as integral components of galactic nuclei,
possibly playing a decisive role in shaping the structure and
morphology, as well as the gas and thus stellar content of massive galaxies. The
\LCDM paradigm for structure formation together with a number of
surprisingly tight relations between the SMBH masses and fundamental
properties of the galactic bulges hosting them -- e.g. the spheroid
luminosity \citep{1995ARA&A..33..581K} and mass
\citep{1998AJ....115.2285M, 2004ApJ...604L..89H},
and the stellar velocity dispersion \citep{2000ApJ...539L..13G,
  2000ApJ...539L...9F, 2002ApJ...574..740T} -- suggests co-evolution
of the hierarchically growing galaxies and their central black holes
(e.g. \citealp{1969Natur.223..690L, 1998A&A...331L...1S,
  2000MNRAS.311..576K}).

At present, variants of particle-based Smoothed Particle Hydrodynamics
(SPH, e.g. \citealp{2004NewA....9..137W, Springel2005MNRAS}),
grid-based Adaptive Mesh Refinement (AMR, e.g. \citealp{1997ApJS..111...73K,
  2002A&A...385..337T}) codes, or moving mesh codes 
\citep{2010MNRAS.401..791S}, are the methods of choice for
numerical simulations of cosmological galaxy formation, exploiting the high
dynamic range and spatial flexibility in resolution. The gravity
solvers in these codes either employ a particle-mesh scheme or are
tree based and assume the simulated system to be collisionless, with
two-body relaxation time-scales exceeding 
the age of the system. In 'tree' algorithms \citep[][see also
\citealp{1993ApJ...414..200M} for a collisional tree
code]{BarnesHut1986Natur} the gravitational force on a single particle
from a distant group of particles is approximated by a multipole
expansion about the group's centre of mass, and the N-body particles
represent massive tracer particles that sample the
underlying smooth gravitational potentials. To reduce the graininess
of the potential, gravitational forces are 'softened' on small spatial
scales and the softening length $\varepsilon$ is the natural
resolution limit of the code. Unfortunately, this also means that
close two-body encounters with a massive body such as a supermassive
black hole can - by construction - not be computed accurately. An
alternative, much more cost-intensive approach of 
calculating the gravitational forces in an astrophysical system is the
direct summation of each particle's gravity on every other particle in
the system. Combined with high-order integrators, this method is a very
accurate way to calculate the gravitational forces and is widely used to simulate
collisional N-body systems \citep[e.g.][]{1999PASP..111.1333A,
  2003gnbs.book.....A,2010ascl.soft10076H}.

Owing to the inherent limitations in the numerical methods, studies of
stellar dynamics in the vicinity of SMBHs to date 
could only probe single separate aspects of the full problem.
On the one hand, direct N-body simulations on SMBH binary 
dynamics in the centres of isolated galaxies or merger remnants
(e.g. \citealp{1991Natur.354..212E, Milosavljevic&Merritt2001ApJ,
  2003ApJ...596..860M, 2006ApJ...642L..21B, 2007ApJ...671...53M,
  2011ApJ...732...89K, 2012ApJ...749..147K, 2011ApJ...732L..26P}) use
idealized initial conditions to represent the inner parts of galaxies
in which the massive binaries are then embedded and evolved. Due to the 
steep scaling of required computing time with particle number of 
order $\mathcal{O}(N^2)$ these studies are still rather limited in
the particle number, hindering a self-consistent
treatment of full galactic environments. On the other hand,
simulations of galaxy mergers or cosmological simulations of structure
formation including SMBHs
(e.g. \citealp{SpringelDiMatteoHernquist2005MNRAS,
  2008ApJ...676...33D, JohanssonEtAl2009ApJ, 2009MNRAS.398...53B,
  2012MNRAS.420.2859M, 2012ApJ...754..125C, 2014arXiv1408.6842S}) have
difficulties to capture the dynamics of SMBHs and their surrounding stars below
the resolution limit. This leads to uncertainties, e.g. in the
dynamical friction time-scales for the SMBHs, and affects the density-
and velocity-profiles of the stellar background through interactions
with the SMBHs, and the hardening and merging time-scales of close
SMBH pairs. The latter is generally assumed 'a priori' to happen fast
in these simulations, much reducing the accuracy and predictive
power of such simulations.
Hence,
there is still substantial uncertainty in the current understanding of
the dynamical evolution of SMBHs 
and their surrounding star clusters in realistic cosmological
settings, which directly feeds back 
into uncertainties in our understanding of how SMBH singles or
multiples influence the structure of galaxies. 

The main goal of this paper is to help remedy these shortcomings by
combining the best parts of the two numerical approaches: a
regularization method to efficiently {\em and} accurately compute the
dynamics close to the black holes and a fast tree code to treat the
global galactic dynamics. This goes in line with the development of
similar recent hybrid codes (as discussed in the next Section) and a
software interface designed to efficiently combine 
different stand-alone code architectures \citep{2009NewA...14..369P,
  2013CoPhC.183..456P}. With the new algorithm, we will be in a
position to better take into account the relevant dynamical processes
regarding the interaction between SMBHs and the stars in their
environments and other (SM)BHs, in principal without limitations on
the spatial and temporal resolution down to scales where other types
of physical phenomena become important, e.g. gravitational wave
induced coalescence of SMBH binaries or the tidal disruption of
low-angular-momentum stars \citep[e.g.][]{2005PhRvL..95l1101P,
  2009MNRAS.392..332L}. This might be an important next step
towards investigating the dynamical co-evolution of SMBHs and their
host galaxy nuclei in a self-consistent manner in galaxy-scale or
cosmological simulations.

In this paper, we present the details of our hybrid N-body code
which will help us to focus on the role of gravitational dynamics in the
interplay between supermassive black holes and realistic
representations of the central regions of their host galaxies.
The paper is structured as follows. In
Section \ref{Hybrid}, we describe the details and structure of the new
hybrid code ``rVINE'' and test its performance in Section
\ref{Performance} after we have described the numerical set-up in
Section \ref{Setup}. First tests on the code in comparison with
the direct N-body code NBODY7 and the tree code VINE are presented in
Section \ref{Compare}. We discuss our results in Section
\ref{Discussion} and, finally, summarize and draw our conclusions
in Section \ref{Conclusions}.\\ 

%
\begin{figure*}
  \begin{center}
    \includegraphics[width=\textwidth]{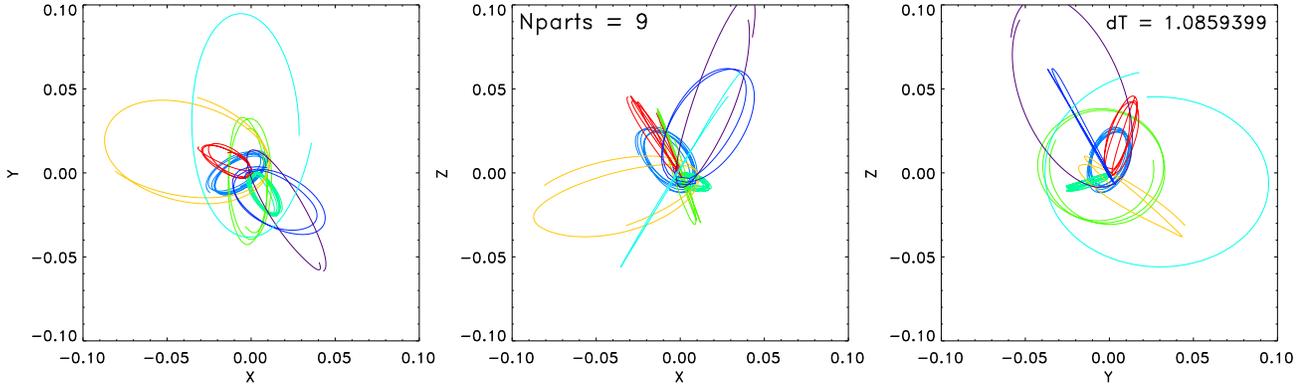}
  \end{center}
  \caption{Projected trajectories of regularized particles in the
  chain around a supermassive black hole. The trajectories are shown
  in a time interval $\Delta t \approx 1$, which corresponds to $\sim$
  one crossing time of the subsystem shown.}
  \label{pic:orbits}
\end{figure*}

\section{A new regularized tree code}
\label{Hybrid}

In this Section we present the structure of the regularized
tree code called ``rVINE''.
A number of currently available N-body
codes employ regularization techniques intended for the integration of
strong gravitational interactions but are primarily developed for
integrations of collisional systems such as star clusters or
the dense central regions of galactic nuclei
  \citep[see][]{2003Ap&SS.285..367A, 2007MNRAS.378..285A,
    2008MNRAS.389....2H, 2009NewA...14..630G, 2012MNRAS.424..545N,
    2013hpc..conf...52B, 2015arXiv150403687W}. In addition, there 
  are two recent hybrid codes combining tree and N-body codes. The
  BRIDGE code \citep{2007PASJ...59.1095F} uses a simple fixed
  time-step oct-tree, while the {\tt Bonsai} tree code
  \citep{2012JCoPh.231.2825B} is an oct-tree run entirely on
  GPUs. Both hybrids use a symplectic mapping method
  \citep{1991AJ....102.1528W} to couple the tree to a
  direct N-body algorithm with a fourth-order Hermite integrator.

In rVINE we compute the evolution
of a sub-system of particles near the black hole by means of a
regularization method, while regions of the galaxy further out with
long relaxation times and basically unaffected by the 
presence of the black hole, are integrated by a collisionless
tree code. To this end, we combine two published algorithms: a version
of the tree/SPH code VINE \citep{2009ApJS..184..298W,
  2009ApJS..184..326N} and the algorithmic regularization (AR) chain 
method \citep[][see also \citealp{2010CeMDA.106..143H} for a
general discussion]{2002CeMDA..84..343M}. The latter was kindly provided as
a stand-alone code by Seppo Mikkola. 

VINE is an OpenMP-parallelized tree/SPH code employing a binary tree
algorithm and an individual hierarchical block time-step
scheme \citep{2009ApJS..184..298W, 2009ApJS..184..326N}\footnote{Note
  that in the present paper we will discuss new developments
  done in the parts of VINE related with the leapfrog integrator, an
  individual time-step scheme and no Smoothed Particle
  Hydrodynamics.}. 
The AR-chain method is an efficient and extremely accurate method to study
close dynamical few-body encounters and is capable of handling even
(repeated) two-body collisions \citep[][see also 
\citealp{1999AJ....118.2532P, 1999MNRAS.310..745M,
  1999CeMDA..74..287M}]{2002CeMDA..84..343M}. 
This is achieved by effectively removing any singular behaviour in the
equations of motion by a time transformation in the Hamiltonian of the
regularized sub-system \citep{2002CeMDA..84..343M}\footnote{However,
  the singularities do formally remain in the transformed equations of
  motion --  unlike in schemes based on
  \citet[][(KS)]{kustaanheimo1965perturbation} regularization, which
  applies a time {\em and} a coordinate transformation.}. The
coordinates and the (original) time and their respective 'momenta' are
integrated using a simple leapfrog method. In addition, a Bulirsch-Stoer
extrapolation method \citep{Gragg65, BS66} is applied to guarantee
high accuracy, 
as well as a chain concept of smallest inter-particle vectors to
reduce round-off errors \citep{1990CeMDA..47..375M,
  1993CeMDA..57..439M}. In our present 
version of the AR-chain, chain particles are sorted according to their
gravitational forces, not according to their distance. It also
includes a method to handle velocity-dependent forces, which allows
us, in principle, to treat additional viscous and relativistic terms
in the regularized force calculations \citep{2006MNRAS.372..219M}. The
new code, however, is purely Newtonian at the present stage.

There exist other regularization schemes which are comparable to the
AR-chain in accuracy and speed, e.g. the KS-wheelspoke
\citep{1974CeMec..10..207Z, 2007MNRAS.378..285A} and the KS-chain
method \citep{1993CeMDA..57..439M}. Due to limitations of these
methods in the context of stellar dynamics around one or several
SMBHs we decided to use the AR-chain as our principal
regularization algorithm. For example, large mass ratios may lead to a
loss in numerical accuracy for the less massive bodies in a KS-chain,
whereas the wheelspoke has difficulties in treating multiple heavy
bodies on an equal footing.

%
\begin{figure*}
\vspace{-1.5cm}
\begin{center}
  \includegraphics[width=0.5\textwidth,angle=270]{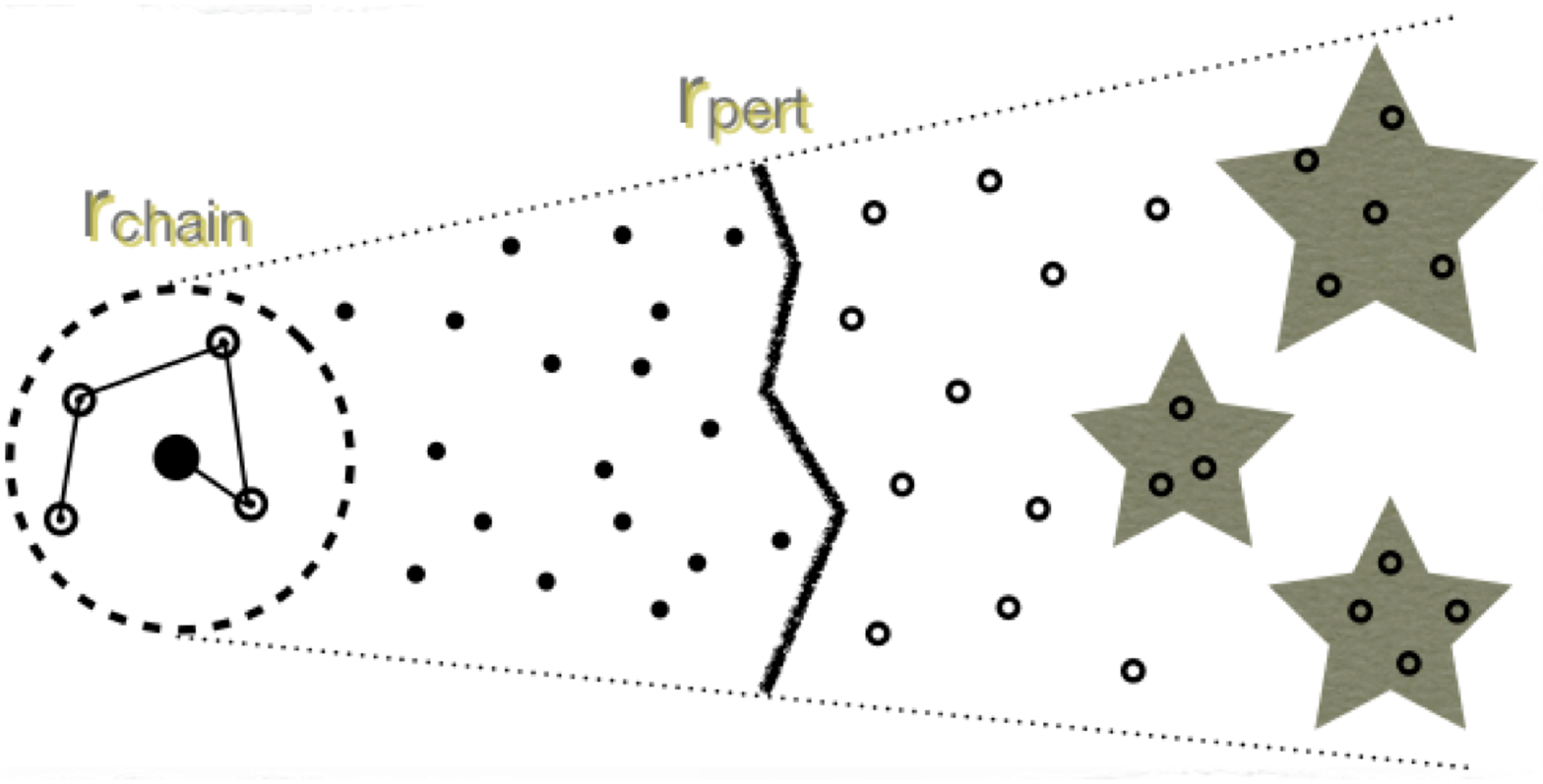}
\end{center}
\vspace{-1.5cm}
\caption{Illustration of the different integration regions near 
  a regularized massive particle in the hybrid AR-regularized tree
  code. The centre-of-mass reference body, consisting of the members
  within the regularized subsystem ($r<r_{\rm chain}$), is surrounded
  by a swarm of nearby particles ({\it 'perturbers'}) which are
  considered as external force terms in the chain force calculations
  ($r<r_{\rm pert}$, where the critical radius $r_{\rm pert}$ depends
  on a tidal criterion) and, themselves, experience direct N-body
  forces from the resolved chain (see text). Further out, we indicate
  the regime where the direct integration of particles (open circles)
  switches to the multipole approximations, depending on the
  acceptance criteria in the tree code (indicated by the groups of
  particles in the large star symbols).}
\label{pic:IntRegions}
\end{figure*}

The gravitational forces for the majority of the particles are
computed with VINE's fast binary tree scheme without 
regularization, using a pre-defined spline or Plummer 
softening, while particles near a designated massive particle (SMBH)
become members of a compact subsystem which is integrated in the AR-chain 
in its centre-of-mass reference frame.\\
Chain integration starts if any particle comes closer to the SMBH
than $r_{j,\mathrm{BH}} < r_{\rm chain,0}$, where $r_{\rm chain,0}$
defines the initial chain size and is an input parameter which has to
be chosen at the start of a simulation as described in the following.

The intended purpose for including particles in the chain integration
is twofold. Firstly, we want to accurately follow close orbits near
the black holes, i.e. within a fair fraction of the SMBHs' influence
radii, and secondly, we need to overcome the limitations posed by the
gravitational softening, i.e. for encounters within $\sim$ a few times
$\varepsilon \equiv \max{(\varepsilon_{\rm
    BH},\varepsilon_\star)}$, where $\varepsilon_\star$ and
$\varepsilon_{\rm BH}$ denote the gravitational softening lengths of
the stellar particles and SMBHs, respectively. Hence, we determine the
initial chain radius, 
\begin{equation}
  \label{eq:Rchain0}
  r_{\rm chain,0} = \max(\alpha \cdot r_{\rm infl}, \beta \cdot \varepsilon),
\end{equation}
at the start of the chain, where $\alpha$ and $\beta$ are input parameters
and $r_{\rm infl}$ is the gravitational influence radius of the
SMBH. For practical purposes, we define $r_{\rm infl} = \min(r_{\rm
  infl}^{\rm 2M},r_{\rm infl}^{\rm \sigma})$ using two commonly used
proxies for the gravitational influence radius of the SMBH, being 1)
the radius enclosing twice the mass of the SMBH, $r_{\rm infl}^{\rm 2M}
= r(<2\,M_{\rm BH})$, and 2) the radius within which the gravitational
force of the SMBH dominates over the self-gravity of the stellar
background, $r_{\rm infl}^{\rm \sigma} = \mathrm{G}M_{\rm BH} /
\sigma^2$. Here, $M_{\rm BH}$ is the mass of the black hole and G the 
gravitational constant. The velocity dispersion $\sigma$ is determined
by averaging over the nearest $50$ particles outside the chain. We
generally find good results for $1 \le \alpha \le 1.5$ and $1 \le
\beta \le 2$.

The chain's center-of-mass is treated as a massive particle in
the tree\footnote{Henceforth we will call this particle simply the
  ``chain particle'', denoting the chain's centre-of-mass particle
  that is advanced in the tree code; not to be confused with a single 
  ``particle in the chain'', which we will equally call a ``chain
  member'' from now on.}, i.e. it is included in the tree force
calculations and advanced in time within the tree code. The chain
particle is advanced in time on the smallest tree time-step and we
formally set the gravitational softening of the chain particle to zero
in the tree-code. 
Tree particles that become members of the chain are converted into 'ghost'
particles with no further advancement in the tree. This is done by
assigning a very small, but finite-sized mass in the tree code data
structure, rendering their contribution to the gravitational forces
negligible. Furthermore, ghost particles are not allowed to determine
the size of the time-step in the tree. If the chain is active, the
member particles within the chain are advanced using the AR-chain
integration, every time particles in the tree code on the smallest
time-step level are being advanced.

The equations of motion of the chain members include
external forces exerted by a set of nearby tree particles we call
``perturbers'', which are identified via a tidal criterion,
\begin{equation}
  \label{eq:rcrit}
  r_{j,\mathrm{CoM}} < \left( \frac{2}{\gamma_{\rm crit}}\,
    \frac{m_j}{M_{\rm chain}} \right)^{\frac{1}{3}}\times r_{\rm crit},
\end{equation}
where $m_j$ is the mass of perturber $j$, $M_{\rm chain}$ is the
total mass in the chain, and $\gamma_{\rm crit}$ a dimensionless
parameter which defines the relative tidal perturbation on the
chain. We typically set $r_{\rm crit} = \min(r_{\rm chain}, r_{\rm
  chain,0})$ in the simulations presented here.
For better accuracy, the perturber positions relative to the center-of-mass 
particle are predicted (to 1st order) to the current time at each
force calculation in the AR-chain. The perturber forces have to be
predicted many times during the numerous sub-step cycles of the
Bulirsch-Stoer extrapolations. To improve the overall performance of
the chain part of the code we, therefore, have implemented parallel
routines for the predictions of the perturbers and the computation of
the perturber forces.

%
\begin{table*}
\caption{Parameters for the different N-body calculations.}
\label{Tab1:ICs}      
\begin{minipage}{\textwidth}     
\centering          
\begin{tabular}{ c c c c c c c c c c c c}
\hline
\hline                             
Simulation\footnote{
All quantities are given in code units.}$^, $\footnote{Note that,
throughout the text, added suffixes will state the actual particle
numbers of the simulations.} & $N_\star$ & $M_{\rm tot}$ & $r_0$ 
& $m_\star/M$ & $\varepsilon_\star$ & $N_{\rm BH}$ & $r_{\rm BH}^{\rm
  init}$ & $v_{\rm BH}^{\rm init}$ & $M_{\rm BH}/M_{\rm tot}$ &
$\varepsilon_{\rm BH}$ \\ 
\hline                             
A\_Nbody7 & 100k & 1.0 & 1.0 & $10^{-5}$ & -- & 1 & 2.14 & 0.46 &
$10^{-3}$ & -- \\
A\_Vine-E1/A\_Gadget-E1 & 100k & 1.0 & 1.0 & $10^{-5}$ & 0.02 & 1 & 2.14 & 0.46
& $10^{-3}$ & 0.1 \\
A\_Vine-E2/A\_Gadget-E2 & 100k & 1.0 & 1.0 & $10^{-5}$ & 0.02 & 1 & 2.14 & 0.46 &
$10^{-3}$ & 0.02 \\
A\_rVine & 100k & 1.0 & 1.0 & $10^{-5}$ & 0.02 & 1 & 2.14 & 0.46 & $10^{-3}$ & -- \\
\hline
B\_Nbody7 & 10k-100k & 1.0 & 1.0 & $10^{-4}-10^{-5}$ & -- & 2 & 0.1/0 &
0.28/0 & $5 \times 10^{-3}$ & -- \\
B\_Vine & 10k-1M & 1.0 & 1.0 & $10^{-4}-10^{-6}$ & 0.01 & 2 & 0.1/0 &
0.28/0 & $5 \times 10^{-3}$ & 0.01 \\
B\_rVine & 10k-1M & 1.0 & 1.0 & $10^{-4}-10^{-6}$ & 0.01 & 2 & 0.1/0 & 0.28/0 & $5 \times 10^{-3}$ & 0.01 \\
\hline
\hline
\end{tabular}
\end{minipage}
\end{table*}
%
%
\begin{figure*}
  \begin{center}
    \includegraphics[width=\textwidth]{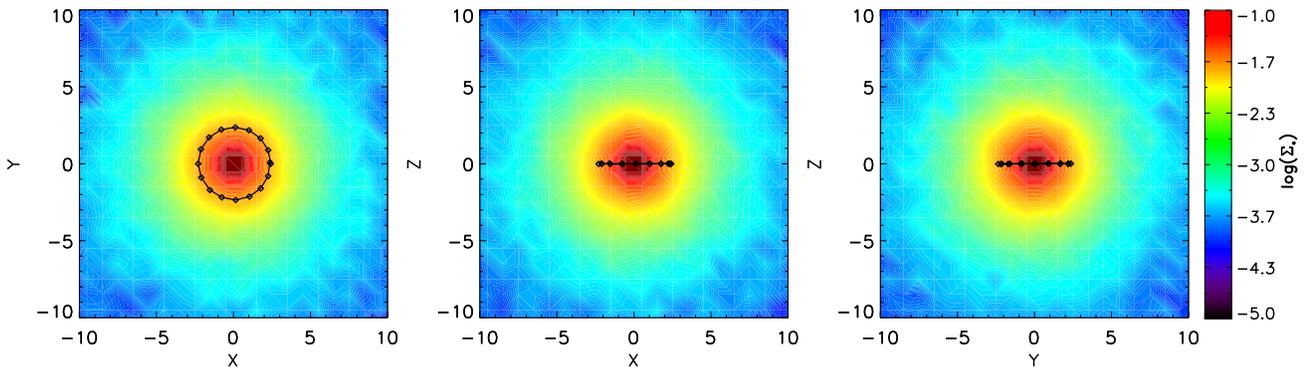}
  \end{center}
  \caption{Projections of the initial stellar surface densities (in
    code units) for a realisation of model A\_rVine\_100k.
    The orbital evolution of a SMBH, placed on a circular orbit at the half-mass
    radius, is shown as the black solid line and symbols for the first
    orbital period.}
  \label{pic:ProjectHernquist}
\end{figure*}
%
%
\begin{table}
\caption{Parameters for the rVINE calculations.}
\label{Tab2:rVINE}      
\begin{minipage}{0.5\textwidth}     
\centering          
\begin{tabular}{ c c c c c}
\hline
\hline                             
Simulation\footnote{Note that, throughout the text, added suffixes
  will state the actual particle numbers of the simulations.} &
$\alpha$ & $\beta$ & $\gamma$ & $\gamma_{\rm crit}$ \\
\hline                             
A\_rVine & 1.0 & 1.0 & 1.5 & $10^{-4}$ \\
B\_rVine & $0.1 - 0.3$ & $ 1.0 - 4.0$  & 1.5 & $10^{-4}/10^{-5}$ \\
\hline
\hline
\end{tabular}
\end{minipage}
\end{table}

Likewise, the force contributions of individual ('resolved') chain
particles are added to the gravitational forces for the perturber
particles during the force updates in the tree. The gravitational
force on the chain particle is also corrected by resolving
the chain particles. If the chain particle is active, we first
calculate the gravitational force exerted on it by particles within
the tree code, but subtract the tree forces (i.e. direct, softened N-body 
interactions) from the perturber particles again. Then the
(direct) gravitational force from the perturbers on each {\it
  individual} chain particle is calculated, and added up as a
correction to the chain particle's force according to   

\begin{equation}
  \label{eq:CoMforce}
  \vec{a}_{\rm CoM,corr} = \frac{1}{M_{\rm chain}}\sum_{i=1}^{N_{\rm chain}} m_{{\rm chain}, i}\,\vec{a}_{{\rm chain},i}.
\end{equation}

After advancing the chain one full time-step its membership is
updated.
Perturber particles are added to the chain if they come closer than
the ``chain radius'', i.e. 
\begin{equation}
  \label{eq:Absorb}
   r_{j,\mathrm{CoM}} < r_{\rm chain},
\end{equation}
where we define the chain radius as the largest distance of
any chain member relative to the chain centre-of-mass in the last
chain step. Particles are removed from the AR-chain if they recede far
enough from the chain's center-of-mass  
\begin{equation}
  \label{eq:Escape}
   r_{j,\mathrm{CoM}} > \gamma \times r_{\rm chain,0} \equiv r_{\rm escape}
\end{equation}
and move radially away from the chain centre. We typically choose
$\gamma = 1.5$. Upon removal of a chain member, its mass, position and
velocities in the global (i.e. tree) reference frame are restored. If
the number of chain members would fall below $N_{\rm chain} < 2$ after an
escape the chain is terminated and the remaining two chain members are
restored in the tree. Likewise, the chain is terminated in the
(unlikely) case that the designated massive particle is removed from
the chain --- unless the chain membership includes several massive
particles, in which case one of the remaining SMBHs is
chosen to be the new centre-of-mass particle in the tree-code. A
representative example of near-Keplerian, perturbed orbits of
particles in the chain is illustrated in Figure \ref{pic:orbits}.

The regularization of particle orbits near a
designated (massive) particle, leads to 
different integration regions for different particles 
in the code, as illustrated in
Figure \ref{pic:IntRegions}. With increasing 
distance to the chain's centre-of-mass the particles are either 
\begin{enumerate}
\item regularized members of the chain (within the dashed
  circle in Figure \ref{pic:IntRegions}), feeling external gravitational forces
  from the perturbers (filled circles) only,  
\item perturber particles which feel the gravitational forces of the
  individual chain members, or  
\item tree particles, whose gravitational forces
onto the chain particle may either be calculated via
direct summation (open circles) or approximated by a multipole expansion
of a group of particles (indicated by star symbols), depending on the
values for the acceptance criteria chosen in the tree.
\end{enumerate}
For cost reasons, we restrict the total number of chain
members and perturbers. We have tested up to values of
$N^{\rm max}_{ch} = 250$ and $N^{\rm max}_{pert} = 5000$ without any
loss in the stability of the code.

In short, a typical time-step in the hybrid code proceeds as
follows.\\
(1) Determine the next time-step and active particles for the tree.\\
(2) If the chain is not active, or, the system does not evolve on the
smallest time-step, continue with step (6).\\
(3) Integrate the members of the chain using the AR-chain
method. Treat the external forces exerted by nearby particles
(``perturbers'') as perturbations to the members' gravitational
forces.\\
(4) Check for absorption by and escape from the chain. Perform
  a search to identify the perturbers via a tidal criterion in regular
  time intervals (see Equation \ref{eq:rcrit}).\\
(5) Terminate the chain if $N_{\rm chain} < 2$.\\
(6) Perform regular leapfrog time-step in the tree code in the
'drift-kick-drift' (DKD) scheme:\\
(a) Update tree particle positions at the half time-step.\\
(b) Compute gravitational forces for the active tree particles. If
the chain particle or any perturber is active include force
corrections due to the gravitational forces between the perturber
particles and the resolved members of the chain.\\
(c) Update tree particle positions and velocities at the full
time-step.\\
(7) Update tree particle time-steps in the individual time-step scheme
and update the tree structure if necessary.\\
(8) If the chain is not active check for particles near the SMBH
that fulfill the conditions to begin chain regularization.\\

\section{Numerical set-up}
\label{Setup}

In this Section we shortly describe the numerical set-up and the
different numerical models we will introduce in the following
Sections. In particular, we will detail the way the code works by
discussing an example simulation and testing its performance
for a number of different code parameters in Section
\ref{Performance}. In Section \ref{Compare} we test the code
against a number of other currently available N-body codes, such as
the VINE and Gadget-3 tree codes and the NBODY7 direct summation code.
All of the rVINE, VINE, or Gadget-3 simulations presented in this
paper were run on the COSMOS cluster at DAMTP, Cambridge. For the
NBODY7 simulations we used single nodes on the Wilkes cluster at the
High Performance Computing (HPC) Service of the University of Cambridge,
consisting of a Dell PowerEdge T620 server \'a 12 Intel Xeon E5-2670
CPUs plus two NVIDIA K20 GPUs.

As our principal numerical model we use a non-rotating
\citet{Hernquist1990ApJ} sphere to represent the galactic 
nucleus. The Hernquist density profile follows a $\rho(r) \propto r^{-1}$
power-law at small radii ($r << r_0$) while converging quickly, with
$\rho(r) \propto r^{-4}$, to a finite mass at large radii ($r >>
r_0$). All models are set-up with unit total mass ($M_{\rm tot}=1$) and
scale radius ($r_0=1$), and the gravitational constant $G$ is set to 
unity in all codes used throughout this paper. Note that, for
convenience, we will use a system of code units in the plots shown
throughout the paper. Identifying our model, for instance, with a
small spherical (dE/S0) galaxy or a galactic
nucleus of mass $M_{\rm tot}=10^{10}\Msun$ and scale length $r_0 = 1
\kpc$ yields time units and velocity units of $\sim 4.7 \Myr$ and 
  $\sim 207 \kms$, respectively. In code units the half-mass radius 
of the Hernquist sphere is given as $r_{\rm 1/2} = (1 + \sqrt{2})\,r_0
\approx 2.41$ with a half-mass dynamical time of $t_{\rm dyn,h} =
27$. In the different simulations considered here the 
galaxy is realised with total particle numbers in the range
$10^4-10^6$ of equal-mass stellar particles ($m_\star \propto
N^{-1}$) for VINE and rVINE.\footnote{Note also that we plan to employ much
  higher particle numbers in rVINE, well above what we have used here
  for the comparison tests, in future (see Section \ref{Discussion}).}
For NBODY7, however, we restrict ourselves to $N\le 10^5$ particles due to
the steep $\mathcal{O}(N^2)$ scaling of the direct N-body code.
Additionally, we  place a few massive particles representing the
SMBHs in the simulation, with initial conditions depending on the
specific simulation run. Models using a single SMBH are denoted as
models 'A' (Sections \ref{Performance} and \ref{DynF}) while models
using two SMBHs are denoted as models 'B' (Sections \ref{Performance}
and \ref{BinSMBH}). To reduce the scatter in our simulations, we will
show the results for all models as averages
over several independent Monte-Carlo realisations of initial
conditions with different random seeds. Depending on $N$, we set up
four different realisations for $20$k $\le N \le 100$k, two realisations for $100$k
$< N < 1$M and one realisation for $N=1$M. 
 Details of the set-up
together with the explicit name for each simulation run are given in
Table \ref{Tab1:ICs}. As an example, we show the early orbital
evolution of a SMBH on the underlying initial stellar surface density 
for a particular realisation of models 'A\_rVine\_100k' in Figure
\ref{pic:ProjectHernquist} (see also Section \ref{DynF}).

%
\begin{figure}
  \begin{center}
    \includegraphics[width=\columnwidth]{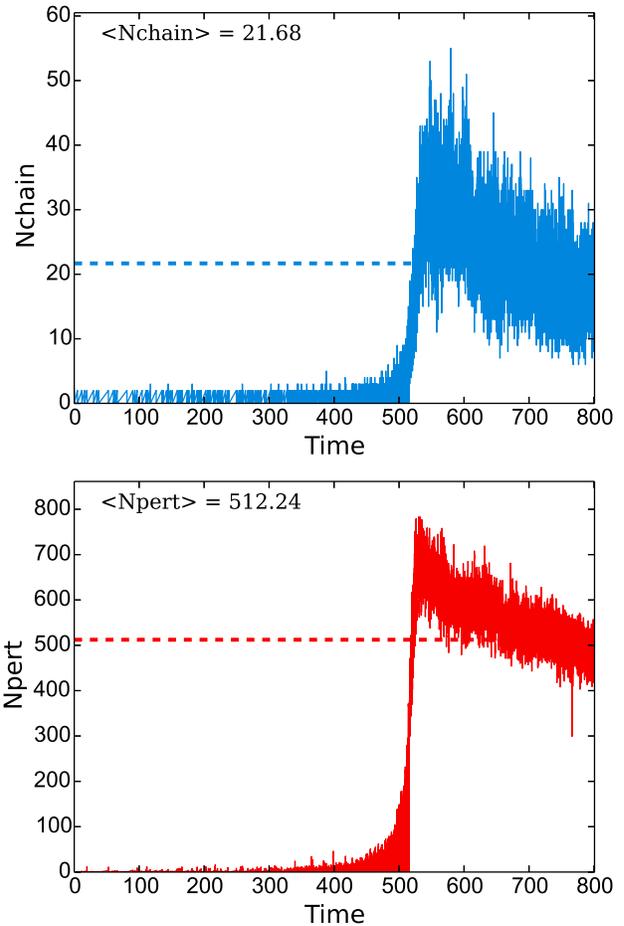}
  \end{center}
  \caption{Time evolution of the number of members in the chain and the
    number of perturbers in a realisation of model A\_rVine\_100k,
    with one SMBH initially set on a circular orbit (see Section
    \ref{DynF}). Average numbers are given as text and indicated by
    the dashed lines.} 
  \label{pic:NchainVsTime}
\end{figure}

In all simulations using either VINE or rVINE, we introduce a
simple Plummer softening \citep{1963MNRAS.126..223A} with different
values for the star particles $\varepsilon_\star$ and the SMBHs
($\varepsilon_{\rm BH}$). However, interactions between chain members
as well as interactions with the (active) chain centre-of-mass
particle in the tree code are not softened at all. For the force
calculations in the tree 
we use a ``Gadget''  multipole acceptance criterion (MAC) with error
tolerance $\theta = 10^{-3}$ and conservative values for the accuracy
parameters $\tau_{\rm acc, \star} = \tau_{\rm vel, \star} = 0.3$ and
$\tau_{\rm acc, BH} = \tau_{\rm vel, BH} = 0.003$ for the leapfrog
integration (see \citealp{2009ApJS..184..298W}). In the Gadget-3
simulations the same integration accuracy, $\tau = 0.02$, is adopted
for both stellar and BH particles, along with an error tolerance of
$5\times 10^{-3}$.  For NBODY7 we set the chain radius to $1.25 \times
10^{-3}$ and do not make use of softened gravitational
interactions. The total accumulated relative energy error stayed below
$\lesssim 10^{-4}$ for all simulations with an accuracy parameter of
$\tau_0 = 0.02$. The parameter ranges used to define the perturber
particles and chain members (see Section \ref{Hybrid}) in the rVINE
simulations can be found in Table \ref{Tab2:rVINE} .

\section{Testing the code and code performance}
\label{Performance}

In this section, we carry out some basic examples and tests on
the code performance of the AR-regularized tree code. 

In Figure \ref{pic:NchainVsTime} we illustrate the time evolution of
the number of particles in the chain (upper panel) and the number of
perturbers (lower panel) for one realisation of model A\_rVine\_100k,
following the orbital  
evolution of a SMBH in a Hernquist sphere consisting of $N=10^5$
particles (see also Section \ref{DynF}).
The SMBH is assigned a mass of $M_{\rm BH} = 10^{-3}\, M_{\rm tot}$
and set on an initially circular orbit at the half-mass
radius ($r_{\rm BH}(t=0) = 2.41$). As the SMBH sinks  
to the centre the number of chain members as well as perturbers
increases significantly at time $t \sim 520$, i.e. at about the time
the SMBH sinks to the central high-density regions within $\lesssim
10$\% of the half-mass radius. Shortly after we reach
the maximum number of 55 particles in the chain and a maximum number of
$\sim 780$ perturbers, while the average numbers are only $\langle
N_{\rm chain} \rangle \sim 20$ and $ \langle N_{\rm  pert} \rangle
\sim 500$, respectively. This highlights the need to perform some
exploratory simulations to determine whether rVINE can actually handle
the range of particle densities around the SMBH encountered over the
simulated time span.
 
%
\begin{figure}
  \begin{center}
    \includegraphics[width=0.95\columnwidth]{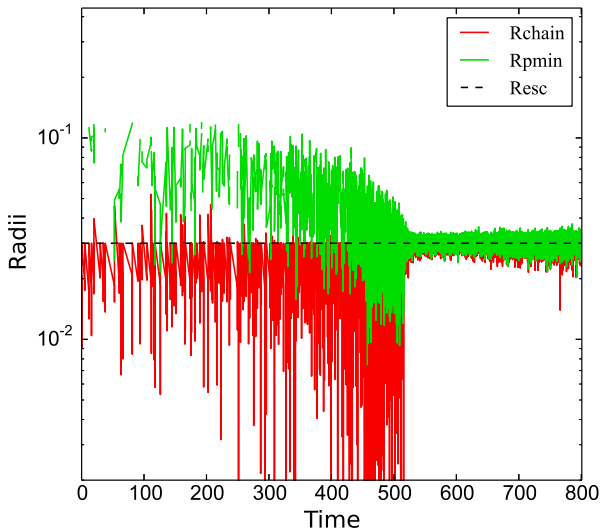}
  \end{center}
  \caption{Time evolution of characteristic radii in a realisation of
    model A\_rVine\_100k, with one SMBH set on a circular orbit (see
    Section \ref{DynF}): the radius of the chain (red solid line), the
    distance of the nearest perturber to the chain centre-of-mass
    (green solid line) and the radius of escaping chain particles 
    (black dashed line).}
  \label{pic:RadVsTime}
\end{figure}

Overall, a fraction of $6.9 \%$ of the $10^5$ particles, was subject
to integration in the chain during this run. Chain integration was
active for a total of $42 \%$ of the total run time, $t_{\rm
  max}=800$. At the early stages of the simulation, the chain is used
only intermittently to treat the occasional strong stellar encounters
near the SMBH in the low density environment. At later phases, when
the SMBH is near to the centre, the chain is used more intensively,
with the chain being active continuously for $\sim 285$ time
units.  Stellar particles are typically included repeatedly in the
chain with 
an average number of $\sim 8$ recurrences.

Figure \ref{pic:RadVsTime} shows the corresponding time evolution of
some characteristic radii from the simulation shown in Figure
\ref{pic:NchainVsTime}. The escape radius (black dashed line), given
by Equation \eqref{eq:Escape}, serves as an effective upper bound for
the chain radius (red solid line). For times $t \lesssim 300$ the
minimum disturber distance, $r_{\rm p, min}$, (green solid line) is
generally well above the escape radius, before some perturbers may come
closer to a chain that has only a few and, by chance, quite compact
members while the SMBH is sinking to denser central regions ($300
\lesssim t \lesssim 520$). Once the SMBH is close to the centre of the
Hernquist sphere ($t\gtrsim 550$) and the number of particles in the
chain has increased by a factor of ten, $r_{\rm chain}$ and $r_{\rm p,
  min}$ both oscillate around $r_{\rm escape}$, which naturally arises
when a number of particles lives close to the conditions for both
absorption and escape being satisfied at a certain time. In this case,
in rVINE we prioritise the absorption of near perturbers over a
(delayed) escape of a chain member, until the chain radius has grown
by $5 \%$ over the nominal escape radius. A noticeable oscillation
around the escape radius can then occur if a series of subsequent
absorptions of perturbers near the SMBH takes place. 

%
\begin{figure}
  \begin{center}
    \includegraphics[height=0.9\textheight]{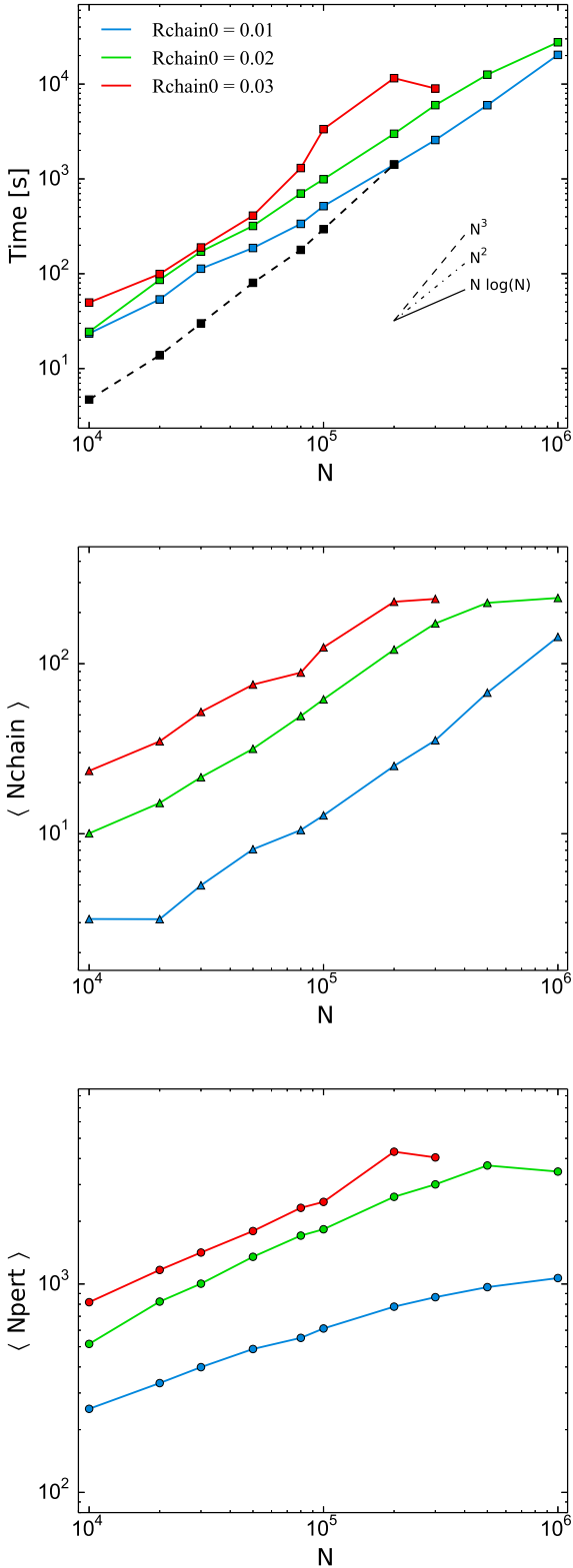}
  \end{center}
  \caption{Performance characteristics for the rVINE code as a
    function of varying initial chain radii in calculations of two SMBHs
    at the centre of a Hernquist sphere, one being set on a circular
    orbit at $x_{\rm BH1} = 0.1$ and the other one being at rest
    at the origin (see also Section \ref{BinSMBH}). The three panels
    show the code run time in seconds ({\it top}), the average number
    of particles in the chain ({\it middle}), and the average number
    of perturbers ({\it bottom}) versus the total number of stellar
    particles in the simulations. All runs were evolved for
    one code time unit ($\Delta{t} = 1$).} 
  \label{pic:CodePerf}
\end{figure}
%

%
\begin{figure}
  \begin{center}
    \includegraphics[width=0.98\columnwidth]{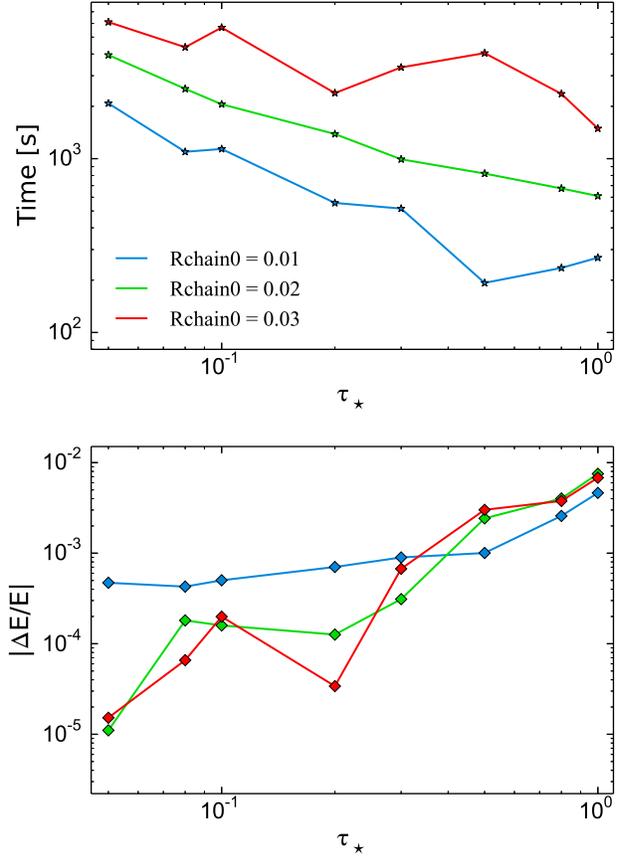}
  \end{center}
  \caption{Performance characteristics for the rVINE code as a
    function of varying initial chain radii for the model used in
    Figure \ref{pic:CodePerf}. Shown are the code run time ({\it top
      panel}) and the relative energy error ({\it bottom panel})
    versus the accuracy parameter of the tree code time-step criteria,
    $\tau_\star$. All runs were evolved for one code time unit
    ($\Delta{t} = 1$).}
  \label{pic:CodePerf2}
\end{figure}

In Figures \ref{pic:CodePerf} and \ref{pic:CodePerf2} we
investigate the performance characteristics of rVINE using a set of
simulations of our basic Hernquist model with two SMBHs,
each having a mass of $M_{\rm BH} = 5 \times 10^{-3} \, M_{\rm
  tot}$. One of the SMBHs is initially set on a circular orbit close
to the centre ($x_{\rm BH1} = \pm 0.1$ and $v_{\rm y, BH1} \approx
0.28$), the other one is at rest at the origin (models B\_Vine and
B\_rVine; see also Section \ref{BinSMBH}). All runs shown in
Figures \ref{pic:CodePerf} and \ref{pic:CodePerf2} were  
performed on a single node (8 CPUs) of Cosmos2, except for a few
comparison runs using NBODY7 (see Figure \ref{pic:CodePerf}, upper
panel) which were performed using 8 CPUs plus acceleration from two
NVIDIA K20 GPUs on the Wilkes cluster.

The three panels in Figure \ref{pic:CodePerf}
show the wall clock time required to evolve the simulation to $t=1$
(upper panel), as well as the average number of particles in the
chain (middle panel) and the average number of perturber
particles (bottom panel) as a function of the initial chain
radius $r_{\rm chain,0}$ and particle number $N$. Interestingly, the
run time does not seem to depend strongly 
on the choice of the initial chain radius; it typically changes by a factor of
$\sim$ a few, and at most by a factor of $\sim 8$, when varying
$r_{\rm chain,0}$ up to a factor of 3. In addition, the scaling with
$N$ is still relatively shallow for all initial chain radii as the
computing time is dominated by the tree. For comparison, we show the
scaling obtained with NBODY7 using the same initial conditions for $N
\le 2 \times 10^5$ (black dashed line). Due to the fact that we can
use the additional acceleration of the two NVIDIA K20 GPUs (plus some
contribution from running on a different system) the total computing
time is actually a factor of $\sim5$ to $\sim10$ times lower
compared to rVINE at the lowest particle numbers. However, the
steeper scaling of NBODY7 ($T_{\rm wall} \propto N^{1.9}$) yields comparable
computing times already for $N \gtrsim 2 \times 10^5$. Extrapolating this
scaling would give clear advantages to rVINE for $N\gtrsim 10^6$
particles in that particular case.

The average number of particles in the chain shows a
roughly linear scaling with total particle number for $N > 2 \times 10^4$ while the
average number of perturbers has a shallower N-dependence. The latter
can be understood by considering the fact that the region of the
perturber particles actually becomes 
smaller with increasing $N$, since the mass of the perturbers scales
as $\propto N^{-1}$ (see Equation \eqref{eq:rcrit}). 
Hence, if the number of particles in the chain were to scale strictly
linearly with $N$, or in the limit of the SMBH dominating the total mass of the
chain, e.g. for a high SMBH-to-star mass ratio and small $r_{\rm
  chain,0}$, we would expect $\langle N_{\rm pert} \rangle$ to be
largely independent of $N$. The observed shallow increase of $\langle
N_{\rm pert} \rangle$ with $N$ is likely to be caused by the
non-trivial non-linear scaling $\langle N_{\rm chain} \rangle$ observed
in the middle panel.
Both $\langle N_{\rm chain} \rangle$ and $\langle N_{\rm pert}
\rangle$ show a scaling going roughly as $N \propto r_{\rm
  chain,0}^2$, as expected for the central parts of the Hernquist
profile, where $M(r) \propto r^2$ \citep[see][their equation
(3)]{Hernquist1990ApJ}. 

The upper panel of Figure \ref{pic:CodePerf2} shows the total run
time of the simulations as a function of the accuracy parameter for
the tree code time-step criteria, $\tau_\star$. In principle
one is free to choose different values for the different time-step
criteria used in VINE (see Section \ref{Setup} and equations (10) - (12) in
\citealp{2009ApJS..184..298W}), but we decided to adopt one, identical
value of $\tau_\star$ for all criteria and fixed the particle
number at $N=10^5$. The time-step size increases linearly with
$\tau_\star$ leading to an overall decrease in the run time for all
values of $r_{\rm chain,0}$, albeit with some non-negligible
scatter. Within the scatter, the total wall time scales roughly as
$T_{\rm wall} \propto r_{\rm chain,0}$ for fixed $\tau_\star$. In
the lower panel of Figure \ref{pic:CodePerf2}, we show the energy
conservation of the code for a given time-step accuracy. To avoid
spurious energy errors upon start-up, we have measured the energy
errors in the interval from $t=4$ to $t=5$. The calculations
become generally less accurate for larger time-step sizes as
expected. However, especially for values of $\tau_\star \le 0.3$, the
scaling is much steeper for larger values of $r_{\rm chain,0}$. i.e. when
a larger fraction of tree code particles are integrated in the more
accurate chain. For the simulations with $r_{\rm chain,0} = 0.02$
and $r_{\rm chain,0} = 0.03$ the relative energy error scales
roughly as $\propto \tau_\star^2$.

\section{Comparison with analytical estimates and other codes}
\label{Compare}

\subsection{Dynamical friction of a SMBH in a Hernquist sphere}
\label{DynF}
As a first test of rVINE, we investigate the orbital evolution
of a massive particle due to dynamical friction in a spherical
non-rotating Hernquist sphere (model A), comparing results from
different simulations using NBODY7, rVINE, VINE, and Gadget-3. The massive
particle of mass $M_{\rm BH} = 10^{-3} \, M_{\rm tot}$ is set on an
initially circular, co-rotating orbit at the half-mass radius ($r_{\rm
  BH}(t=0) = 2.41$) (see also Section \ref{Setup}).

%
\begin{figure*}
\centering 
\unitlength.97cm
\begin{picture}(14.5,22.)
  \put(0,16.2){\epsfxsize = 14.5cm\epsfbox{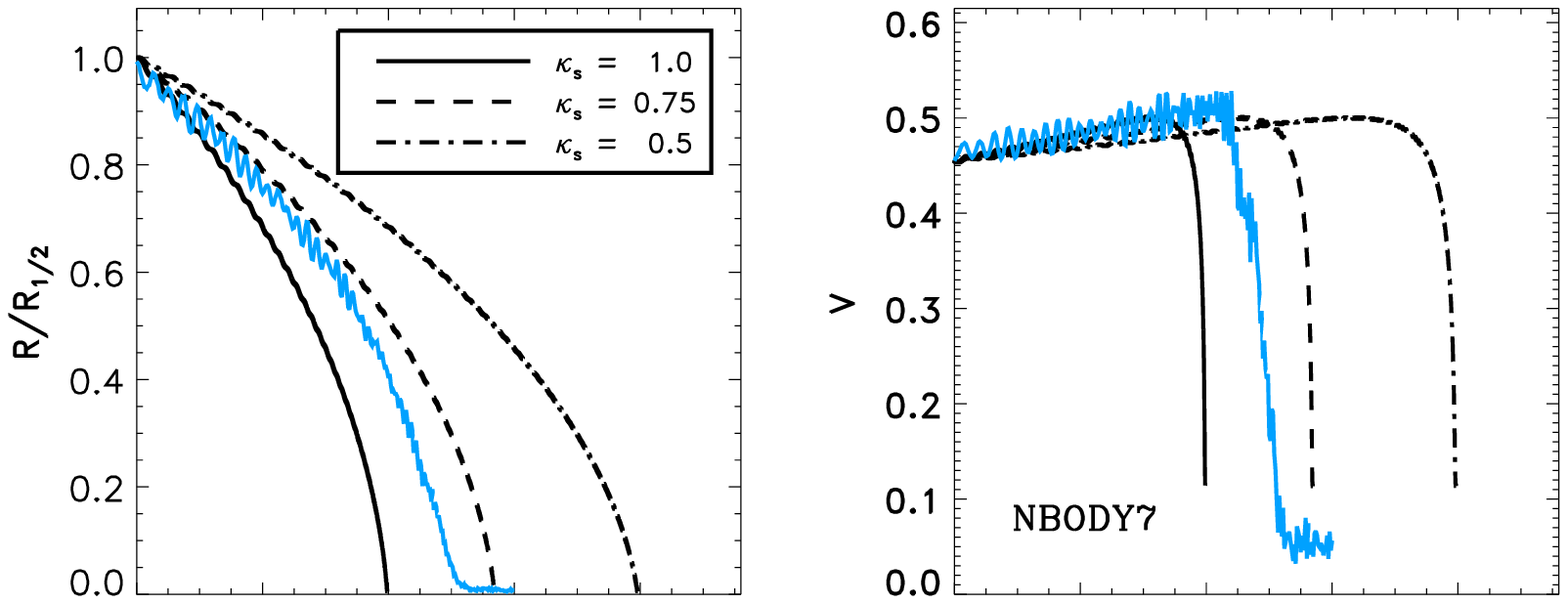}}
  \put(0,10.8){\epsfxsize = 14.5cm\epsfbox{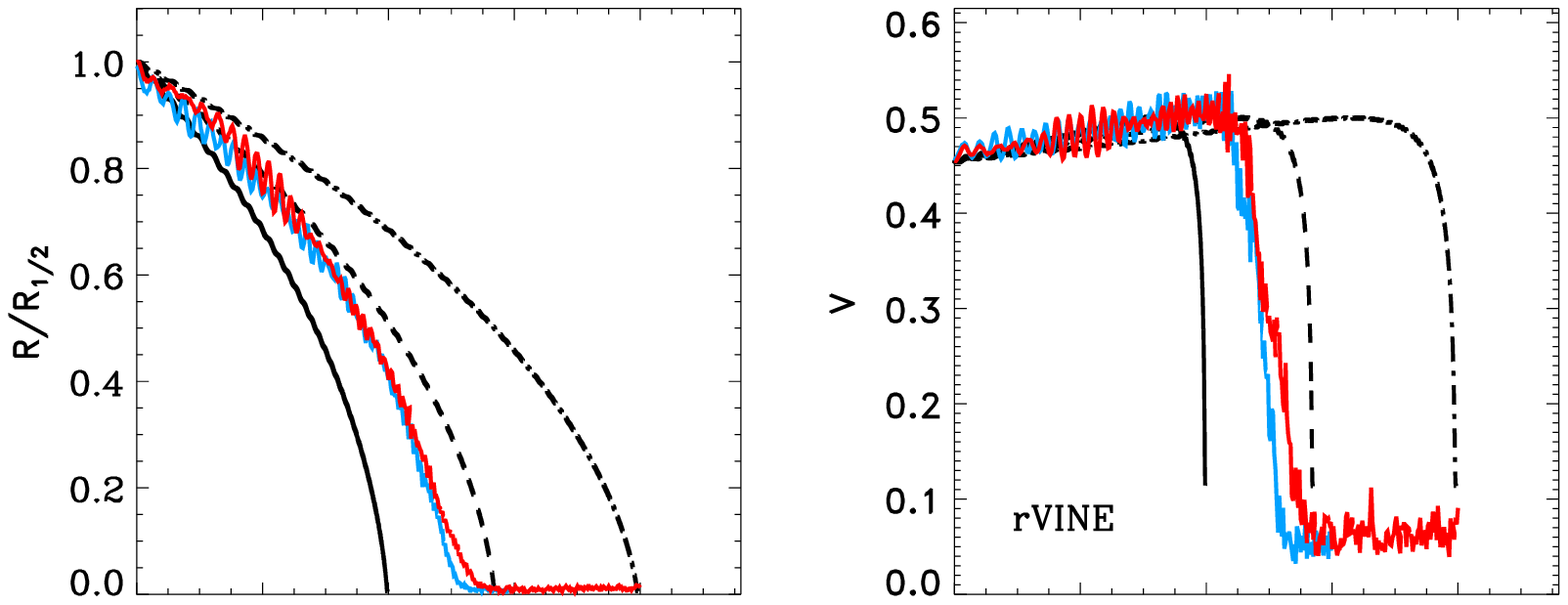}}
  \put(0,5.4){\epsfxsize = 14.5cm\epsfbox{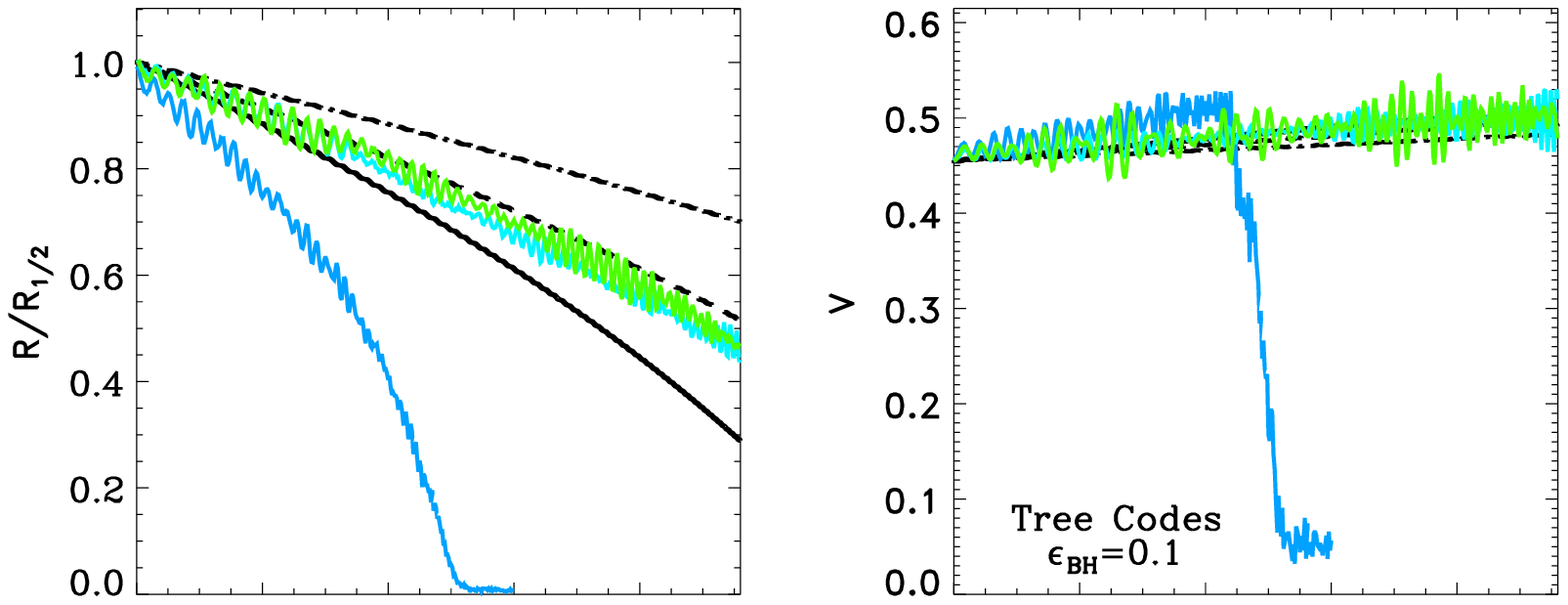}}
  \put(0,0){\epsfxsize = 14.5cm\epsfbox{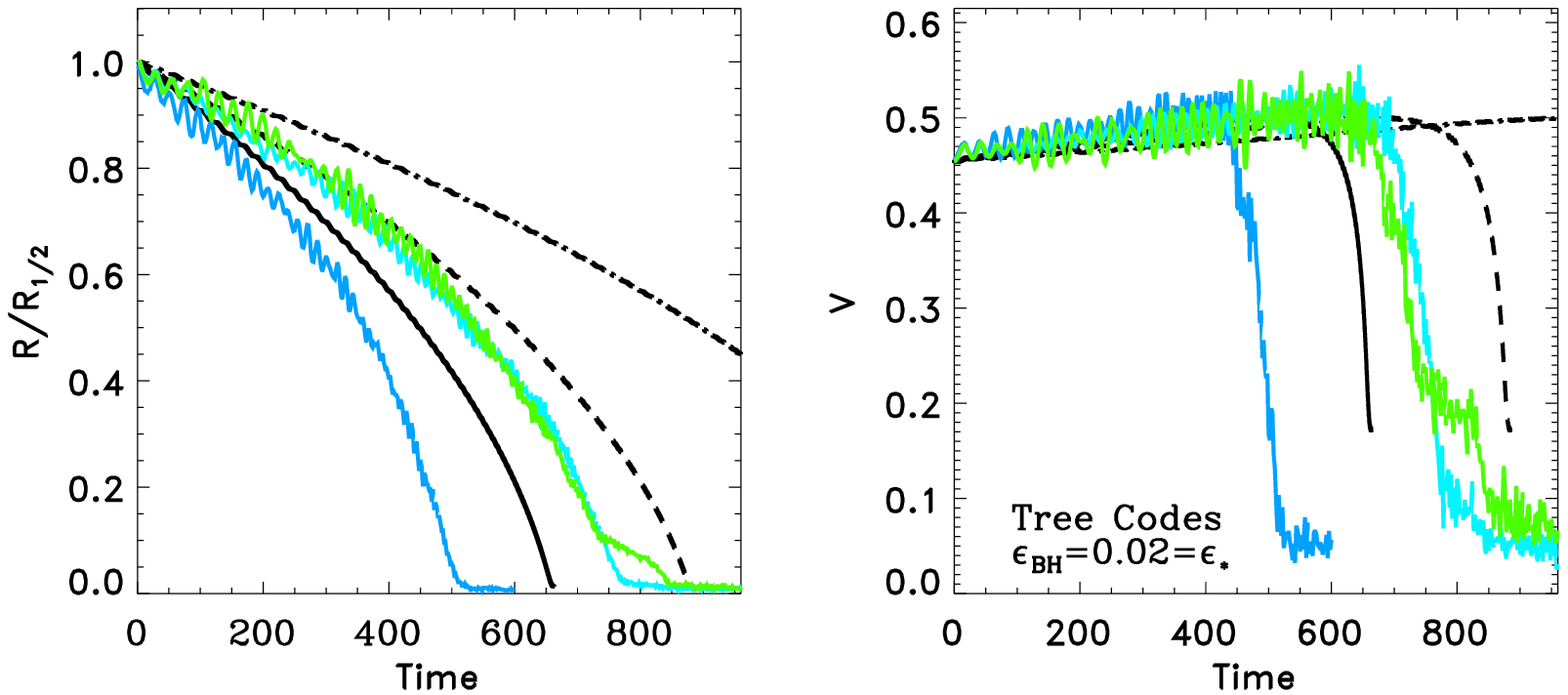}}
\end{picture}
\vspace{-0.25cm}
\caption{Comparison of the distance to the centre ({\it left panels})
  and velocity ({\it right panels}) of a single SMBH, initially set on
  a circular orbit at the half-mass radius of a Hernquist sphere, for
  different code architectures. From top to bottom we show the direct
  summation code NBODY7 (blue; model A\_Nbody7\_100k), the AR-chain
  regularized tree code rVINE (red; model A\_rVine\_100k), and the
  VINE and Gadget-3 tree codes (green and pale blue lines) with
  varying softening lengths, $\varepsilon_{\rm BH} = 0.1$ (models
  A\_Vine-E1\_100k and A\_Gadget-E1\_100k) and $\varepsilon_{\rm BH} =
  0.02 = \varepsilon_\star$ (models A\_Vine-E2\_100k and
  A\_Gadget-E2\_100k). Theoretical expectations from dynamical
  friction theory are given as black lines for different parameters of
  $\kappa_{\rm s}$ ($\kappa_{\rm s}$ parametrizes the
    deviation from a locally isotropic Maxwellian velocity
    distribution, see Section \ref{DynF}). For comparison, we show the
    results for the NBODY7 run also in the other panels. rVINE very well 
    recovers the expected orbital evolution by effectively removing
    the limitations imposed by gravitational softening. Results shown
    are averages over several realisations of the initial conditions,
    as described in the text.}
\label{pic:DF}
\end{figure*}

\subsubsection{Dynamical friction theory}
\label{DynF_theo}
For a meaningful assessment of the different codes, we compare the
simulated SMBH trajectories with theoretical expectations from 
dynamical friction theory \citep{1943ApJ....97..255C,
  2008gady.book.....B}. Assuming a locally isotropic velocity
distribution function, a homogeneous density $\rho$, as well as a
sufficiently large velocity of the SMBH, $v_{\rm BH}$, relative to the
stellar background, the rate of deceleration of the SMBH due to
dynamical friction may be given by the standard formula
\citep[cf.][Eq. 8.5]{2008gady.book.....B} 
\begin{equation}
  \label{Eq:DynF}
  \vec{a}_{\rm DF} = -4\pi\,{\rm G}^2 \frac{M_{\rm BH}}{v^3_{\rm
      BH}}\,\rho(v<v_{\rm BH})\, \ln\Lambda\ \vec{v}_{\rm BH},
\end{equation}
where $\ln\Lambda$ is the Coulomb logarithm. Only the mass density
of stars moving slower than the SMBH, $\rho(v<v_{\rm BH})$,
contributes to the dynamical friction. With gravitational softening
there is a limit on the 
maximum gravitational force that may be exerted in any star-SMBH
encounter through an effective minimum impact parameter ${\rm b}_{\rm
  min}$ \citep[see e.g.][]{1976MNRAS.174..467W, 2011MNRAS.411..653J},
\begin{equation}
  \label{Eq:bmin}
  {\rm b}_{\rm min} = 1.5 \cdot \max(\varepsilon_\star,\varepsilon_{\rm BH}).
\end{equation}
Taking this softening effect into account, we write
the Coulomb logarithm as
\begin{equation}
  \label{Eq:lnLambda}
  \ln \Lambda = \ln \left( \frac{{\rm b}_{\rm max}}{\sqrt{{\rm b}_{\rm
          min}^2 + {\rm b}_{\rm 90}^2}} \right),
\end{equation}
where ${\rm b}_{\rm max}$ is the maximum impact parameter and ${\rm
 b}_{90}$ is the impact parameter for a $90^\circ$ scattering event
of the incident star. The choice for the latter two parameters is
often rather arbitrary, with ${\rm b}_{90}$ depending on the typical
velocity $v_{\rm typ}$ of the stars. The maximum impact parameter
${\rm b}_{\rm max}$ is often taken proportional to the orbital radius
of the SMBH $r_{\rm BH}$. Following \citet{2005A&A...431..861J}, we
identify ${\rm b}_{\rm max}$ with the local scale length,
\begin{equation}
  \label{Eq:bmax}
  {\rm b}_{\rm max} = \frac{\rho}{|\nabla \rho|} = \frac{r_{\rm
      BH}}{3-\eta},\ \ \eta \le 2,
\end{equation}
where the last equation is true for the family of $\eta$-models
\citep{1993MNRAS.265..250D, 1994AJ....107..634T} if $\eta \le
2$. Hence, in the case of the Hernquist profile ($\eta = 2$) we
obtain that the maximum impact parameter exactly equals the orbital
radius of the SMBH, i.e. ${\rm b}_{\rm max} = r_{\rm BH}$. In addition
the $90^\circ$ deflection parameter is given by
\begin{equation}
  \label{Eq:b90}
  {\rm b}_{90} = \frac{{\rm G}\, M_{\rm BH}}{v_{\rm typ}^2} \approx
  \frac{{\rm G}\, M_{\rm BH}}{2\sigma^2 + v_{\rm BH}^2}.
\end{equation} 

Using Equations \eqref{Eq:DynF} - \eqref{Eq:b90}, we evolve the orbit
of a SMBH, initially placed on a circular orbit at the half-mass
radius, in the analytic Hernquist potential including the additional 
drag forces due to dynamical friction with a leapfrog integrator. The
density of stars with velocities smaller than the SMBH is represented
as 
\begin{equation}
  \label{Eq:PhaseDens}
  \rho(v < v_{\rm BH}) = \kappa_{\rm s} \cdot \rho_{\rm s}(v < v_{\rm BH}),
\end{equation}
where $\rho_{\rm s}(v < v_{\rm BH})$ denotes a locally isotropic
Maxwellian velocity distribution given by
\begin{equation}
  \label{Eq:erf}
  \rho_{\rm s}(v < v_{\rm BH}) = \rho(r) \times \left[
    \mathrm{erf}(X) - \frac{2X}{\sqrt{\pi}} \mathrm{e}^{-X^2} \right]
\end{equation}
where $X\equiv v_{\rm BH}/\sqrt{2}\sigma$ with dispersion $\sigma$, and erf$(X)$
is the error function \citep{2008gady.book.....B}.
We use $\kappa_{\rm s}$ as a free parameter to account for the
fact that the velocity distribution of the Hernquist model does not
follow a simple Maxwellian distribution, as often used as an
approximation in dynamical friction calculations. Hence, the locally
isotropic Maxwellian velocity distribution corresponds to $\kappa_{\rm
  s} = 1.0$.

\subsubsection{Results}
\label{DynF_results}

The upper panels of Figure \ref{pic:DF} show the evolution of the radial
decay (left) and the velocity (right) of the SMBH with time for model
A\_Nbody7\_100k. NBODY7 is very well suited to follow the dynamical
friction of the heavy body. We gauge run A\_Nbody7\_100k (blue line)
against theoretical orbits for different values of $\kappa_{\rm s} =
(0.5, 0.75, 1.0)$ shown as black lines and use it subsequently as a
reference for the other simulation codes. Initially the orbital
evolution follows quite closely the analytic prediction with
$\kappa_{\rm s} = 0.75$ (black dotted line).
At smaller central distances ($r \lesssim 0.6\,r_{\rm BH}^{\rm init}$)
the dynamical friction seem to act even more efficiently in the full
N-body run. This leads to a slightly faster orbital evolution such
that the SMBH reaches the centre (defined as $r \lesssim r(M <
2\,M_{\rm BH})$, where dynamical friction ceases to be efficient)
within $t \lesssim 500$ time units. In the second row of Figure
\ref{pic:DF} we compare the efficiency of dynamical friction on the
SMBH in model A\_rVine\_100k (red line) and model A\_Nbody7\_100k
(blue line). We find that the early orbital evolution ($r > 0.5\,
r_{\rm BH}^{\rm init})$ in model A\_rVINE\_100k is nearly
indistinguishable from the one in model A\_Nbody7\_100k such that the
SMBH reaches the centre on a very similar timescale, only about $\sim
8\%$ longer than in model A\_Nbody7\_100k, in very good agreement with
the NBODY7 result.

On the other hand, owing to their inherent resolution limitations, we
expect the tree codes to show a significantly slower radial decay
depending on the adopted gravitational softening, via Equation
\eqref{Eq:bmin}. If the SMBH is treated as a quasi collisionless
particle with a large 
gravitational softening length ($\varepsilon_{\rm BH} = 0.1$, third 
row), it decays to only about half its original distance within the
time span ($t\sim 960$) shown in the Figure \ref{pic:DF} in both model
A\_Vine-E1\_100k (green line) and model A\_Gadget-E1\_100k (pale blue
line). The total time to reach the centre is $t \gtrsim 1350$ time
units for both codes, significantly longer (by a factor of $\gtrsim
2.7$) than with NBODY7. This can be understood as a result of the
reduced frictional force for near encounters. Both codes, however,
follow the analytical prediction with $\kappa_{\rm s} = 0.75$ if we
adopt the minimal impact parameter ${\rm b}_{\rm min} = 1.5 \cdot
\varepsilon_{\rm BH}$ from Equation \eqref{Eq:bmin}. Setting
$\varepsilon_{\rm BH} \equiv \varepsilon_\star = 0.02$ (bottom row)
yields an enhancement in the near encounters, but the decay time to
the centre is still about $50\%$ ($70\%$) longer than in the NBODY7
case for model A\_Vine-E1\_100k (A\_Gadget-E1\_100k). Hence,
even this drastic reduction in the gravitational softening of the SMBH
does not significantly improve the accuracy of the dynamical friction
time-scale. 

While all codes capture the effects of dynamical friction as expected,
if we consider softened gravitational force terms, this highlights the
importance of an accurate treatment of close encounters for the
correct description of dynamical friction in N-body tree codes. From
our analysis we conclude that the AR-regularized tree code removes the
limitations due to the gravitational softening of the SMBH imposed on
present-day tree codes by including nearby particles in the chain.

\subsection{Evolution of a SMBH binary at the centre of  a Hernquist sphere}
\label{BinSMBH}

%
\begin{figure*}
  \begin{center}
    \includegraphics[width=0.9\textwidth]{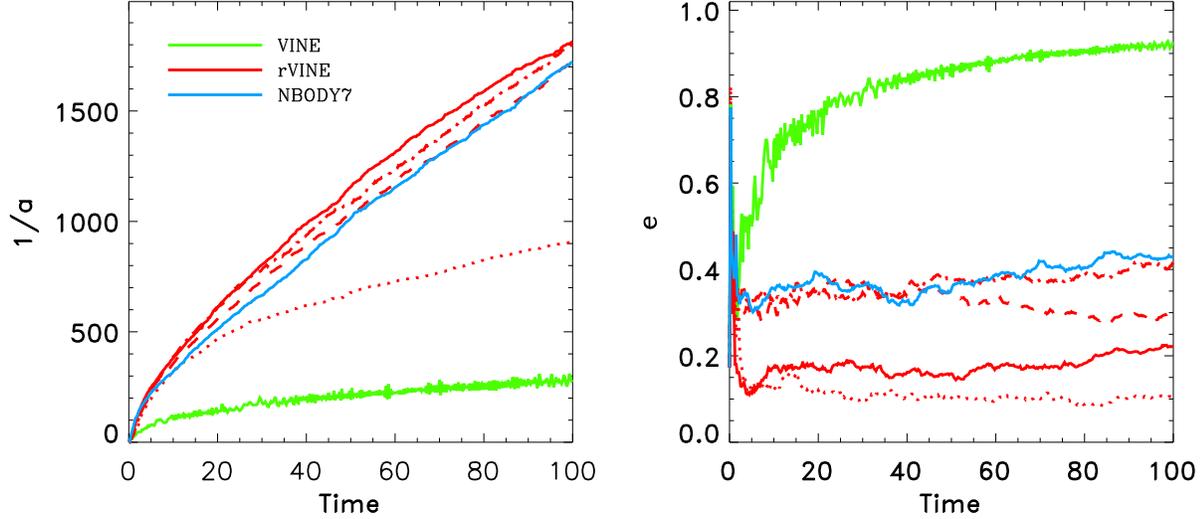}
  \end{center}
  \caption{Evolution of binary parameters of a SMBH binary at the
    centre of a non-rotating Hernquist sphere (model B). Shown are
    binary hardness ($1/a$, {\it left panel}) and eccentricity ({\it
      right panel}) as a function time.
    Different colours indicate simulations using different codes,
    i.e. NBODY7 (blue), VINE (green, with $\varepsilon_{\rm BH} 
    = \varepsilon_\star$), and rVINE (red lines). Different
    line styles for model B\_rVine denote different runs with
    varying initial chain radii ($r_{\rm chain,0} = 0.01$: dotted, $0.02$:
    solid, $0.03$: dashed, and, $0.04$: dot-dashed). The results shown are averages
    over several realisations of the initial conditions, as described
    in the text.}
  \label{pic:BinaryParams}
\end{figure*}

Another crucial test for rVINE is to investigate the hardening
of a SMBH binary, compared to results obtained with NBODY7
and VINE. For our model B, we choose an initial set-up similar to
the one presented in \citet{2007ApJ...671...53M} and
\citet{2014ApJ...785..163V}. In particular, we set up two SMBHs at the
centre of a non-rotating Hernquist sphere with masses $M_{\rm BH} = 5
\times 10^{-3} \, M_{\rm tot}$ and one of the SMBHs on a circular
orbit ($x_{\rm BH1} = 0.1$, $v_{\rm y, BH1} = |v_{\rm circ}| \approx
0.28$) and the other SMBH at rest at the  origin\footnote{Note that in
  \citet{2014ApJ...785..163V} both SMBHs are set on co-rotating
  orbits, while here we set one of the SMBHs at rest at the origin. We
  tested that this different set-up gives similar
  results. Furthermore, in their simulations, the initial velocities are set
  to $|v| = 0.31$, which is about $\sim 10\%$ higher than the circular
  velocity.}. For models B\_Vine and B\_rVine we use gravitational
softening lengths $\varepsilon_\star = 0.01 =\varepsilon_{\rm BH}$
for the stellar and SMBH particles in the tree-code. Particles in the
chain are not softened. The initial distance is chosen slightly larger
than the influence radii of the SMBHs, which typically 
increase in the B\_rVine and B\_Nbody7 runs from $r_{\rm
  infl}^{\rm 2M} \approx 0.1$ to $r_{\rm infl}^{\rm 2M} \approx 0.16$ according
to a decrease in the central density during the simulations. In the
simulations using model B\_rVine, initially, only the co-rotating SMBH is
regularized, but quickly -- for $t \lesssim 1.5$ --- the
second SMBH is captured into the chain. In the B\_Nbody7 runs we use an AR-chain
with a chain radius of $R_{\rm chain} = 1.25\times 10^{-3}$, while we
test different values of the initial chain radius for the AR-chain
in rVINE which we set to a size comparable to the hard binary
semi-major axis, $r_{\rm chain0} \sim \mathrm{a\ few} \times a_{\rm
  hard}$\footnote{Note that employing equally large chain radii would
  not be feasible in NBODY7 without major modifications to the
  code.}. The hard binary semi-major axis 
is given as 
\begin{equation}
  \label{Eq:ahard}
  a_{\rm hard} = \frac{\mu}{M_{\rm BH1}+M_{\rm BH2}} \frac{r_{\rm infl}^{\rm 2M}}{4},
\end{equation}
with $\mu = (1/M_{\rm BH1} + 1/M_{\rm BH2})^{-1}$ being the reduced
mass. Initially, $a_{\rm hard} = r_{\rm infl}^{\rm 2M}/16 \approx 6.25
\times 10^{-3}$.

In Figure \ref{pic:BinaryParams} we show the time evolution of the SMBH
binary hardness ($1/a$, left panels) and eccentricity (right panels)
for the first $t=100$ time units in the simulations of model B with
$N=10^5$ particles. 
If we set a large initial chain radius, $r_{\rm chain,0} = 0.02$
(red solid lines), $r_{\rm chain,0} = 0.03$ (red dashed lines), or
$r_{\rm chain,0} = 0.04$ (red dot-dashed lines), model B\_rVine\_100k
agrees well with model B\_Nbody7\_100k (blue lines) 
showing an efficient hardening with a nearly constant hardening rate
$\frac{d}{dt}(1/a)$ for $t \gtrsim 10$, and a rather mild binary
eccentricity ($e<0.4$). On the other hand, when setting the initial
chain radius comparable to $a_{\rm hard}$ and the softening parameter,
$r_{\rm chain,0} = 0.01$ (red dotted lines), the hardening of the SMBH
binary proceeds much slower since we start to miss out on some close 
encounters with the hard binary, owing both to the effect of softening
and the lower cross-section of the chain.  
Model B\_Vine\_100k (green lines), on the other hand, shows a
qualitatively different picture, with a significantly
larger separation between the two SMBHs ($1/a \approx 300) $ and a
high binary eccentricity of $e \lesssim 0.95$ at the end of the
simulations. While the reduced binary hardening is clearly due to both
the softened gravitational forces near the SMBHs and the reduced rate
of loss-cone refilling in the collisionless tree-code, the reason for
the high binary eccentricity is unclear.

%
\begin{figure*}
  \begin{center}    
  \unitlength1.cm
  \begin{picture}(15.3,21.6)
    \put(0,14.4){\epsfxsize = 15.3cm\epsfbox{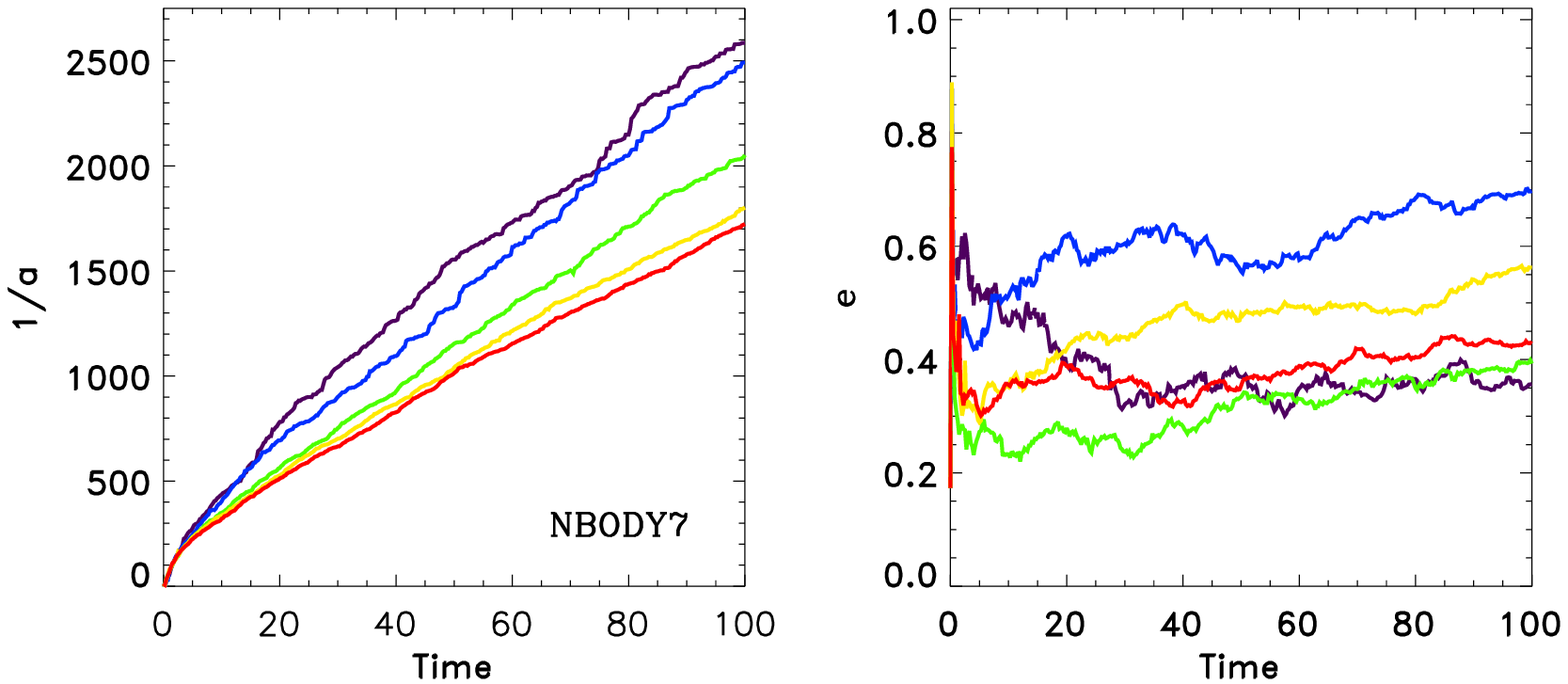}}
    \put(0,7.2){\epsfxsize = 15.3cm\epsfbox{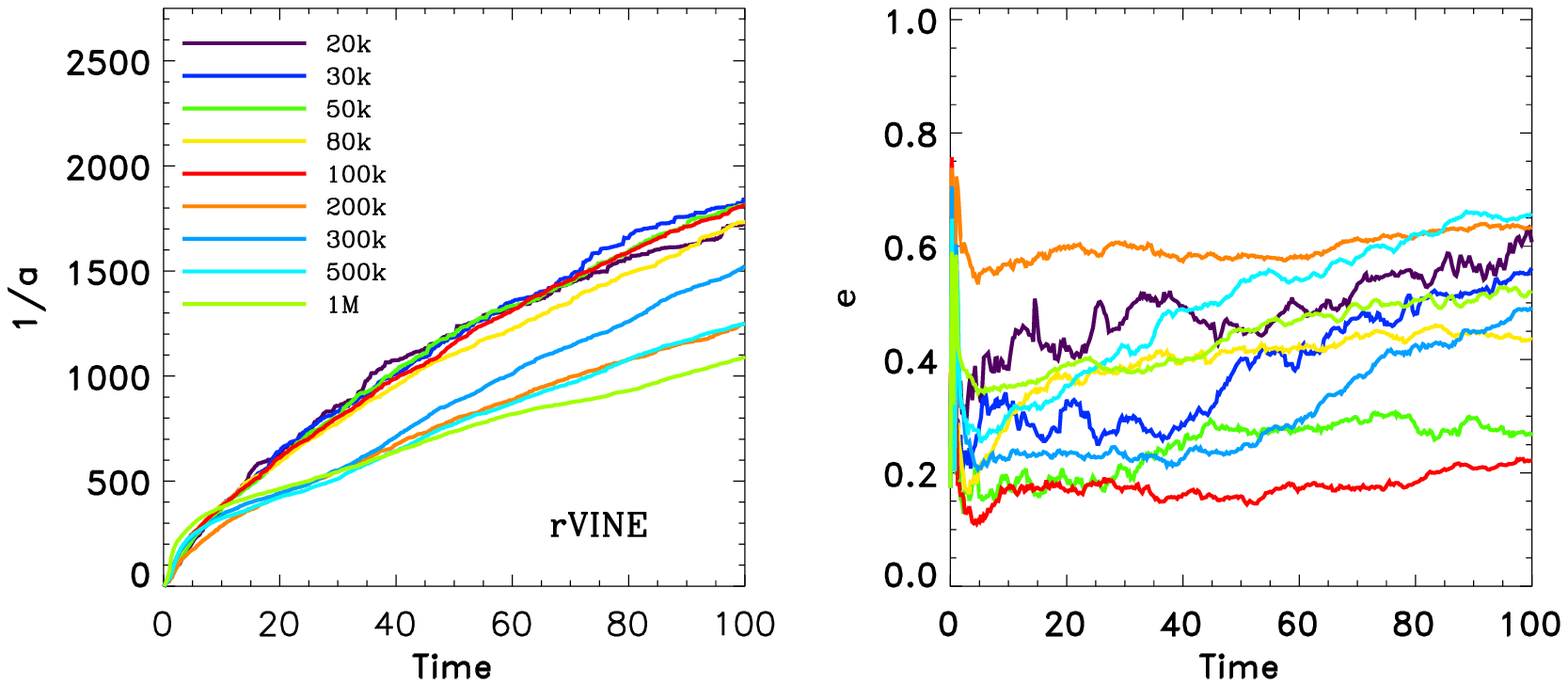}}
    \put(0,0){\epsfxsize = 15.3cm\epsfbox{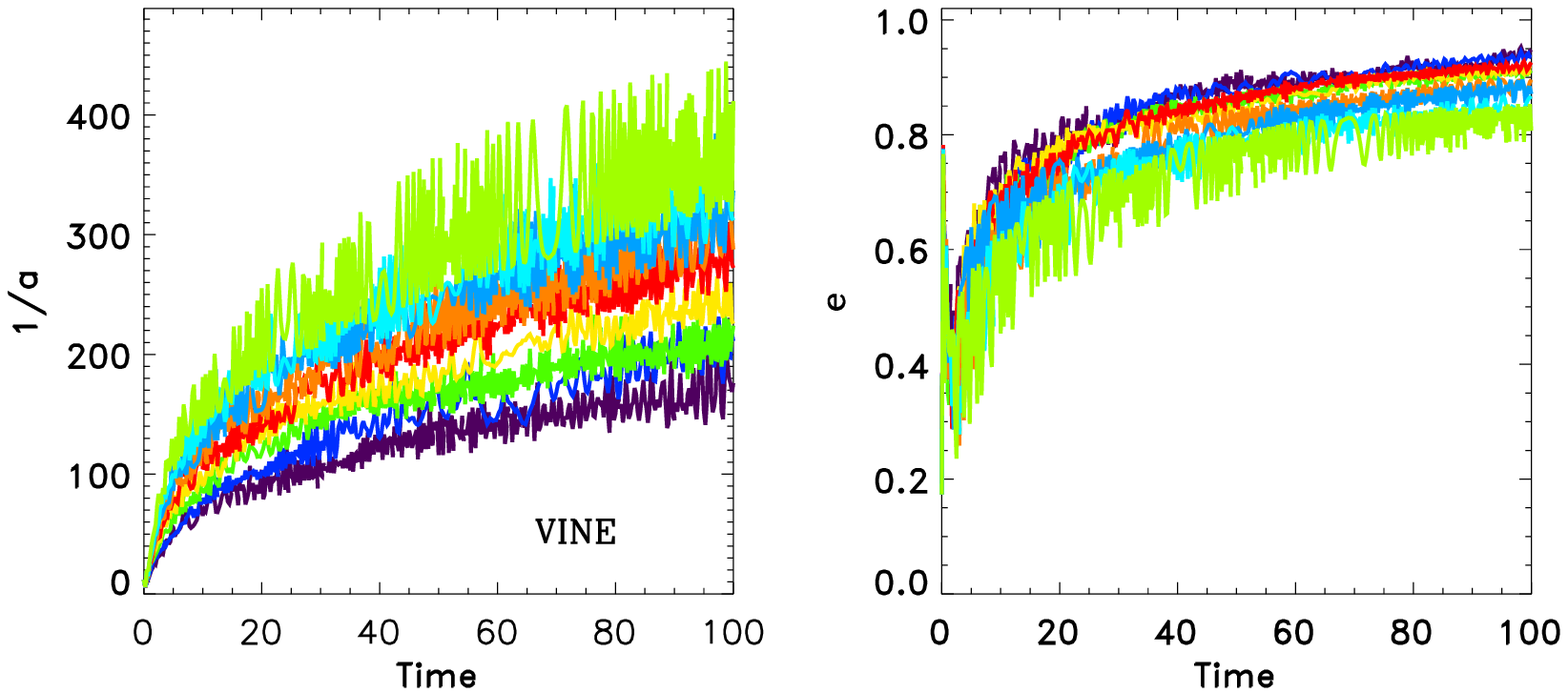}}
  \end{picture}  
  \end{center}
  \caption{Time evolution of binary hardness ($1/a$, {\it left
      panels}) and eccentricity ({\it right panels}) as a function of
    particle number N for the three different codes in model B:
    NBODY7 ({\it top row}), rVINE ({\it middle row}) and VINE
    ({\it bottom row}). Shown are simulations with particle numbers
    increasing from $N=20$k to $N=1M$ from top to bottom, with colours
    indicated in the legend. The initial chain radius is set to
    $r_{\rm chain,0} = 0.02$ in the simulations of model B\_rVine. The
    results shown are averages over several realisations of the
    initial conditions, as described in the text.}
  \label{pic:BinaryParamsAvN}
\end{figure*}

Figure \ref{pic:BinaryParamsAvN} extends the analysis of Figure
\ref{pic:BinaryParams} to a range of different particle numbers. For
 VINE and rVINE we investigate simulations with particle numbers
 ranging from  $N = 2 \times 10^4 - 10^6$ and between $N=2 \times 10^4
 - 10^5$ for NBODY7. In the initial stages 
 of the simulation ($t \lesssim 5$), when the loss-cone is full and
 dynamical friction is still efficient, the binary parameters evolve
 qualitatively similar in all three codes, with a steep rise in $1/a$
 and a quite broad range of moderate eccentricities $0.1 \le e \le
 0.7$). However, thereafter models B\_Vine (bottom row) quickly evolve
 to high eccentricities while the binary semi-major axis typically
 stalls at $a \lesssim 0.5\, a_{\rm hard}$. In models B\_Nbody7 (top row)
 and B\_rVine (middle row), on the other hand, a 
 hard binary with $a \le a_{\rm hard}$ forms quickly within $t\approx
 2.5-5$. Models B\_Nbody7 and B\_rVine again show an overall much
 more efficient binary hardening and eccentricity evolution than in
 B\_Vine. The binary semi major axis and eccentricity reach final
 values of $1/a \lesssim (1100 - 2600)$ and $e\lesssim (0.2-0.7)$ in
 the B\_Nbody7 and B\_rVine runs, respectively, depending on $N$. Both in
 models B\_Nbody7 and B\_rVine the binary hardening decreases with
 increasing $N$. Since this is due to
 the lower efficiency of collisional loss-cone refilling for larger
 $N$, however, the decrease is more pronounced in model B\_Nbody7. As
 expected, in model B\_Vine we again have a very weak evolution of the
 binary hardening ($1/a\lesssim 400$) but high final eccentricities
 ($0.8 \lesssim e \lesssim 0.95$) with a very weak dependence on $N$.

%
\begin{figure}
  \begin{center}
    \includegraphics[width=0.42\textwidth]{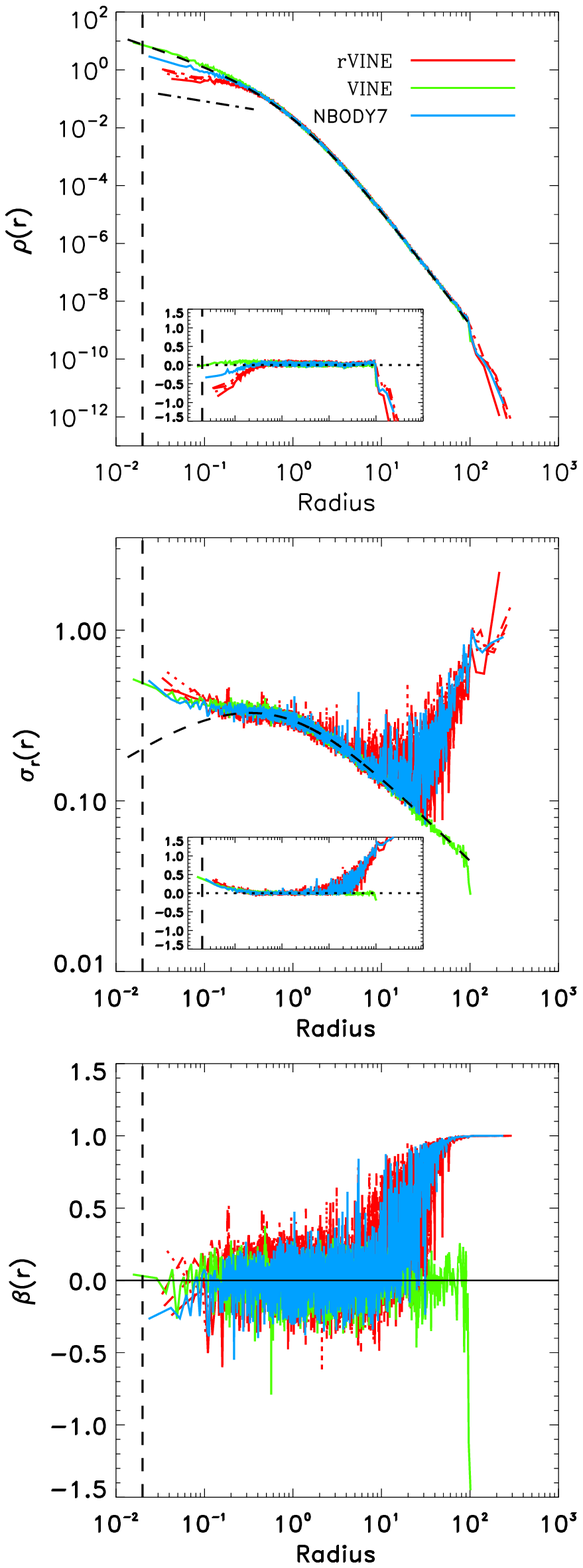}
  \end{center}
  \caption{Radial profiles of the final density ({\it upper panel}), radial velocity
    dispersion ({\it middle panel}) and anisotropy parameter ({\it
      bottom panel}) for the models shown in Figure
    \ref{pic:BinaryParams}: B\_Nbody7\_100k (blue), B\_Vine\_100k
    (green) and B\_rVine\_100k (red lines). For 
    comparison, the initial Hernquist profiles are shown with the
    dashed lines, the dashed-dotted line shows an inner density
    profile following $\rho \propto r^{-0.5}$, and the vertical dashed
    lines indicates the resolution limit in model B\_Vine\_100k, given
    as $\sim 2 \, \epsilon_{\star}$. Inserts show the residuals
    from the initial Hernquist profiles for the same radial
    range. Shown are averages over several realisations of the initial
    conditions.}
  \label{pic:DensProfile}
\end{figure}

In Figure \ref{pic:DensProfile} we show the effect the ejection of low
angular momentum stars at the bottom of the potential has on the
structure of the galactic nucleus. Shown are radial profiles for the
density (upper panel), the radial velocity dispersion (middle
panel), and the velocity dispersion anisotropy parameter,
\begin{equation}
  \label{eq:Beta}
  \beta_\sigma \equiv 1 - \frac{\sigma_\phi^2 +
    \sigma_\theta^2}{2\,\sigma_{\rm r}^2} 
\end{equation}
(bottom panel), for the models shown in Figure \ref{pic:BinaryParams}
at the end of the simulations. Stars are ejected from the galaxy
centre by the hardening binary on orbits with high radial velocities
in the B\_rVine\_100k (red) and B\_Nbody7\_100k (blue) runs. This
leads to a large increase in the radial velocity dispersion at
galacto-centric radii with $r \gtrsim 10\, r_0$ for these simulations
(upper panel), together with a depression in the central density
profile within $r\le r_{\rm infl}^{\rm 2M}$ and some added mass in the
outskirts of the galaxy (middle panel). The central density profile is
converted from an initial Hernquist profile with inner slope of $\sim
{-1}$ (dashed line) to a profile with $\rho \propto r^{-0.5}$ in the
B\_rVine\_100k and B\_Nbody7\_100k runs. The increase in the
radial velocity dispersion is not seen in the tangential velocity
dispersions such that the anisotropy profile is strongly radially
biased for $r \gtrsim 10\, r_0$. We verified that this is
caused only by particles escaping the system after being ejected in
interactions with the central binary in both codes.
For B\_Vine\_100k (green),
however, all radial profiles evolves insignificantly
throughout the simulation. 
rVINE seems on average to be more effective
in removing mass from the centre than NBODY7. Given that the SMBH
binary hardens by roughly the same amount in both codes, this might be
due to the fact that the central density is replenished more
effectively by the high rate of collisional loss-cone refilling with
stars originating from larger radii in the B\_Nbody7\_100k runs.

\begin{figure}
  \begin{center}
    \includegraphics[width=0.95\columnwidth]{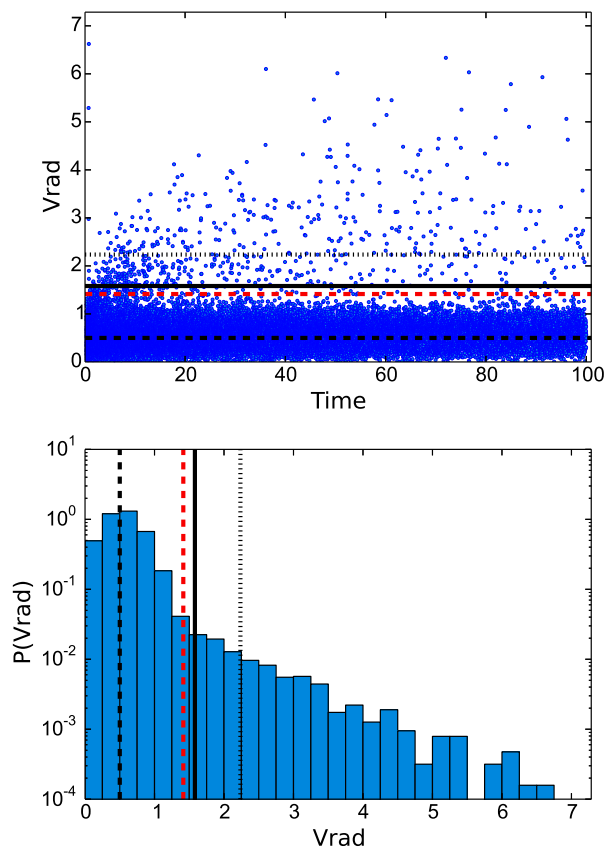}
  \end{center}
  \caption{Radial velocity statistics of particles escaping the
    chain in one of the B\_rVine\_100k simulations, with $r_{\rm
      chain,0} = 0.02$. {\it Top panel:} radial velocities as a
    function of time. {\it bottom panel:} Distribution of radial
    velocities of particles leaving the chain, with the velocities
    binned by $\Delta v = 0.25$. For comparison we show the expected
    kick velocities ($v_{\rm kick} \propto  \sqrt{G\, \mu/a}$,
    \citealp{1974ApJ...190..253S}) for inverse binary semi-major axes
    $1/a = 100$ (dashed black line), $1/a=1000$ (solid black line) and
    $1/a = 2000$ (dotted black line), and the escape velocity from the
    centre of a Hernquist sphere (dashed red line).}
  \label{pic:Vrad}
\end{figure}

We further analyze the properties of the high radial velocity stars in
the realisation using model B\_rVine\_100k shown in Figure
\ref{pic:DensProfile} by examining the radial velocities of stars
escaping the chain in Figure \ref{pic:Vrad}. Not all of the particles 
interacting with the SMBH binary leave the chain on high radial
velocity orbits: the majority of the escapers has $v_{\rm rad} < 1$ at
all times in the simulation. However, for $t \lesssim 10$ there is an
enhanced interaction rate of stars in the initially full loss-cone
with the hardening SMBH binary leading to a significant population of
escapers with radial velocities $1.5 < v_{\rm rad} < 2.5$, comparable
to the expected kick velocities in a slingshot interaction of a field
star with a massive binary with hardness $1/a \sim 1000$. Furthermore
we find a small population of high-velocity outliers ($2 < v_{\rm rad}
< 6.5$) - much higher than the expected kick velocity - which provide
a clear observational signpost of the hard SMBH binary at the centre.

\section{Discussion}
\label{Discussion}

Due to the favourable $\mathcal{O}(N\log(N))$ scaling of the tree
code we should be able to employ (much) higher particle
numbers than in the test calculations presented in this paper in
future applications of the new code. However, as a caveat, we note
that the time spent for the AR-chain calculations scales steeper with
particle number than the time spent for the tree calculations, 
mostly due to the costly extrapolations of the Bulirsch-Stoer method
and the repeated predictions and force evaluations of the
perturbers. 
For the simulations performed for this paper
execution of the chain part of the code has only taken a moderate fraction of 
the total CPU time (typically below $5\%$ with a maximum of
$\sim 20\%$), and with some optimisation it should be possible to
further increase the speed of the code.  
There will, however, be some critical particle number, $N_{\rm crit}$,
at which the AR-chain will become the dominant contributor to the total
computing costs even if only a small fraction of the total particle
number is actually integrated in the chain. 
It is not within the scope of this paper to investigate this in detail
and we leave this to future work where we will use our new hybrid code
for full-scale galaxy simulations.

%

Throughout Section \ref{BinSMBH} we have found a qualitatively similar
evolution of the hardening binary both in NBODY7 and rVINE for chain
sizes of $\sim$ a few times the hard binary distance $a_{\rm hard}$
(Equation \eqref{Eq:ahard}). The agreement is particularly good
for the highest particle numbers studied here ($N \ge 80k$) where
spurious relaxation effects become less important in the direct N-body
code. Similarly, for lower particle numbers ($N \le 50k$), rVINE shows
a shallower N-dependence of the hardening rate since the tree-code
better reproduces the collisionless galactic stellar dynamics at
distances far from the SMBHs.
Hence, the hybrid code seems to catch the relevant dynamical
interactions of a real galaxy better at lower $N$ for our set of
parameters adopted for the chain.  

However, we also note here that the high hardening rates we find in Section
\ref{BinSMBH} in the NBODY7 simulations are somewhat larger than those found
in a number of recent studies of spherical and axisymmetric models
using similar techniques and initial conditions \citep[see
e.g.][]{2013ApJ...773..100K, 2014ApJ...785..163V}. In particular, the
final binary hardness for our B\_Nbody models are on average about a factor
of $\sim 2.5$ higher for comparable particle numbers than the
spherical models studied in \citet{2014ApJ...785..163V}. 
Several reasons could be responsible for this difference including
slightly different choices in the initial positions and velocities of
the SMBHs (see Section \ref{BinSMBH}) or differences in the integration
techniques (parameters for the AR-chain, settings for the
gravitational softening and the integration  
accuracy, etc.), the most likely being the different way of
introducing the SMBHs in the initial conditions. This may lead to an
overestimate of the hardening rate, especially at
the beginning of our simulations.

In Section \ref{DynF} we have found that in the tree codes VINE and
Gadget-3 --- even with the most conservative choice of the SMBH softening
length --- the dynamical friction time-scales for the SMBH to sink to
the galactic centre differ by more than $50 \%$ from the ones in
NBODY7 and rVINE. Hence are present-day cosmological and galaxy merger
simulations suffering from significant (unavoidable) uncertainties in
the SMBH orbital time-scales? Strictly speaking they do, but probably,
in most cases SMBHs are rarely found orbiting 'naked' in their host
galaxies, but are instead embedded in stellar and
gaseous cores or cusps that are then the prime subjects to 
dynamical friction in galaxy interactions. However, it has been shown
that 'naked' BHs might be quite commonly formed after the
disruption of galactic nuclei in gas-rich minor mergers
\citep{2009ApJ...696L..89C, 2014MNRAS.439..474V}, making accurate
estimates of the dynamical friction time-scales of the 
'naked' BHs necessary in these cases.

Of similar importance for the galaxy formation community is
to get better estimates for SMBH binary coalescence time-scales in
order to make robust predictions with respect to the dynamical
evolution of SMBHs in their host galaxies. For
example, studying the exciting possible formation of
systems with multiple SMBHs at high redshift due to the high merger
rate of galaxies relies crucially on (1) an accurate description of
dynamical friction in order to correctly quantify the populations of
binary, triple, etc. SMBHs being present at a given
time, and (2) accurate orbits in order to
reliably calculate the final outcome of the strong multi-body
interactions between the SMBHs in the galactic centres \citep[see
e.g.][]{2012MNRAS.422.1306K, 2011MNRAS.412.2154B}. Obtaining accurate
coalescence time-scales for binary SMBHs in gas-rich galaxy mergers is also
essential for accurate estimates of the 
likelihood of recoiling SMBHs escaping from the rapidly steepening
central potential of the merger remnants
\citep[e.g.][]{2011MNRAS.414.3656S}. This should be particularly 
relevant for large scale cosmological simulations like e.g. 
the recent EAGLE and Illustris \citep{2015MNRAS.446..521S,
  2014MNRAS.444.1518V} simulations that assume fast coalescence of two
SMBHs once their distance falls below the resolution limit \citep[see
e.g.][for details]{2014arXiv1408.6842S}.

\section{Conclusions}
\label{Conclusions}

In this paper we have presented a hybrid code combining an
OpenMP-parallel binary tree code (VINE) with an algorithmic
chain regularization scheme, and report on first tests with the new
code called ``rVINE''\footnote{rVINE is available to anyone
    interested upon request to the authors.}.

We have shown that, using the AR-regularized tree code, we can
significantly improve the numerical accuracy in the calculation of the
gravitational interactions of SMBHs with their close environment. By
comparison with the collision-less code NBODY7, we have verified that
we have overcome some of the fundamental limitations
imposed by the gravitational softening of the SMBHs, as it is used in
traditional tree codes. As a consequence, we are now able to follow
the orbital evolution of SMBHs much more accurately in more
realistic, galaxy-scale settings. We have shown that with the 
new hybrid code we obtain both significantly improved estimates
for dynamical friction time-scales of single SMBHs sinking to the
galactic centres and for the time evolution of
hard SMBH binaries. In particular, using rVINE, we find a clear
N-dependence of the binary hardening rate, a low binary eccentricity
along with a moderate eccentricity evolution, as well as the
conversion of the galaxy's inner density profile from a cusp to a core
via the ejection of low-angular momentum stars on orbits with high
radial velocity, similar to the results obtained with NBODY7 here and
in previous work.

Due to the  modular design of rVine, the AR-chain
part with its hybrid interface should be easily
portable to other codes used for simulations of galaxy formation. 
It will likewise be straightforward to incorporate additional physics
into rVine, e.g. formulations
for hydrodynamics and additional sub-grid models of (1) star formation
and  stellar feedback, and (2) black hole accretion and feedback, or,
the addition of post-Newtonian terms in the AR-chain.
Note also that in the present paper we have restricted ourselves to
the case of regularizing the dynamics of {\em one} single subsystem
only. The next step here is to extend the present code to allow for multiple 
chains in order to handle the regularization of several distant SMBHs
at once. 

Important problems that will benefit from the  accurate dynamical
modelling of the evolution of (binary) SMBHs are 
predictions with regard to SMBH coalescence rates and their associated
gravitational wave background \citep[e.g.][]{1994MNRAS.269..199H}, the
population of SMBHs and AGNs (either living at their host galaxies
centres or being displaced from the central regions by three-body
encounters or gravitational wave recoils), the acceleration of
hyper-velocity stars \citep{1988Natur.331..687H}, and the imprint of
SMBH binaries on the structural and kinematic properties of the inner
parts of the stellar component of galaxies \citep{1991Natur.354..212E, 
Milosavljevic&Merritt2001ApJ, 2013MNRAS.433.2502M}.

Further improved hybrid codes like the one presented here
will help to bridge the gap between different fields of
astrophysics that are currently still separated by huge 
differences in the relevant scales of space, time, and  mass
and should eventually allow to study properly the effects of stellar dynamics
in the sphere of influence of central supermassive black holes 
on the structure of their host galaxies. 

\section*{Acknowledgements}

This work was supported by the Science \& Technology Facilities
Council (STFC) [grant number ST/J001538/1] and the STFC DiRAC project
dp021. We would very much like to thank Seppo Mikkola for providing
the AR-chain routines. SJK is grateful to the hospitality shown by the
Max-Planck-Institut f\"ur Astrophysik in Garching. TN acknowledges
support by the DFG cluster of excellence 'Origin and Structure of the
Universe'. We acknowledge support by Chinese Academy of Sciences
through the Silk Road Project at NAOC (RS, SJA and SJK), through the Chinese
Academy of Sciences Visiting Professorship for Senior International
Scientists, Grant Number $2009S1-5$ (RS), and through the
``Qianren''special foreign experts program of China. The main
numerical part of this work was undertaken using the DiRAC Shared
Memory Processing system at the University of Cambridge, operated by
the COSMOS Project at the Department of Applied Mathematics and
Theoretical Physics on behalf of the STFC DiRAC HPC Facility
(www.dirac.ac.uk). This equipment was funded by BIS National 
E-infrastructure capital grant ST/J005673/1, STFC capital grant
ST/H008586/1, and STFC DiRAC Operations grant ST/K00333X/1. DiRAC is
part of the National E-infrastructure. The NBODY7 part of this work
was realised on the Wilkes GPU cluster at the HPC Service of the
University of Cambridge.

\bibliography{./bib2.2}

\bsp

\label{lastpage}

\end{document}